\documentclass{aa}  
\usepackage{multirow}
\usepackage{fontawesome}
\usepackage{placeins}

\usepackage[colorlinks=true,linkcolor=blue,citecolor=blue,urlcolor=blue]{hyperref}
\usepackage{braket}
\usepackage{graphicx}
\usepackage{txfonts}
\usepackage{ulem}

\newcount\Comments  
\usepackage{color}
\definecolor{darkgreen}{rgb}{0,0.5,0}
\newcommand{\kibitz}[2]{\ifnum\Comments=1\textcolor{#1}{#2}\fi}

\usepackage{mathtools}
\usepackage[absolute]{textpos}
\DeclarePairedDelimiter\abs{\lvert}{\rvert}%
\DeclarePairedDelimiter\norm{\lVert}{\rVert}%

\makeatletter
\let\oldabs\abs
\def\abs{\@ifstar{\oldabs}{\oldabs*}}
\let\oldnorm\norm
\def\norm{\@ifstar{\oldnorm}{\oldnorm*}}
\makeatother

\def\MS{\text{M}_{\odot}}

\def\OC{\omega \: \text{Cen}}

\begin{document} 



\begin{textblock}{5}(11.2,1.65)
\noindent \footnotesize LAPTH-038/24, CERN-TH-2024-131
\end{textblock}

\title{New constraints on the central mass contents of Omega Centauri from combined stellar kinematics and pulsar timing}
\titlerunning{New constraints on the central mass contents of Omega Centauri}

\author{Andr\'es Ba\~nares-Hern\'andez
\inst{1,2}\thanks{\href{mailto:a.banareshernandez@gmail.com}{\texttt{a.banareshernandez@gmail.com}}}
\and
Francesca Calore
\inst{3}
\and
Jorge Martin Camalich
\inst{1,2,4}
\and
Justin I. Read
\inst{5}
}
\institute{Instituto de Astrof\'\i sica de Canarias, La Laguna, Tenerife, E-38200, Spain
\and
Departamento de Astrof\'\i sica, Universidad de La Laguna \and CNRS, Laboratoire d'Annecy-le-Vieux de Physique Théorique, 74940, Annecy, France \and CERN, Theoretical Physics Department, CH-1211 Geneva 23, Switzerland \and Department of Physics, University of Surrey, Guildford, GU2 7XH, United Kingdom}

\authorrunning{Ba\~nares-Hern\'andez et al.}

\date{\today}
 
  \abstract
   {}
{We perform a combined analysis of stellar kinematics with line-of-sight accelerations of millisecond pulsars (MSPs) to probe the mass contents of Omega Centauri ($\OC$). Our mass model includes the stellar mass distribution, a more concentrated mass component linked to the observed MSP population, a generic cluster of stellar remnants (assumed to be more concentrated than the stars and MSPs), and an intermediate-mass black hole (IMBH), allowing us to determine which of these is most favored.}
{We mass-model $\OC$ using the package \texttt{GravSphere} to solve the Jeans equations, including constraints from proper motions, line-of-sight velocities, the surface density profile of the stars, the spatial distribution of MSPs, and the recently measured line-of-sight accelerations of a subset of these MSPs, self-consistently modeling their intrinsic spin-down. We explore the impact of different assumed centers of $\OC$ on our results and we infer the posterior distributions of the model parameters from the combined likelihood using the nested sampling package \texttt{dynesty}.}
{ 
Our analysis favors an extended central mass of $\sim 2 - 3 \times 10^5 \: \MS$ over an IMBH, setting a 3$\sigma$ upper limit on the IMBH mass of $6 \times 10^3\: \MS$. We find that pulsar timing observations are an important additional constraint, favoring a central mass distribution that is $\sim 20\%$ more massive and extended than for models that are constrained by the stellar kinematics alone.
Finally, we find a 3$\sigma$ CL upper bound of
$6\times10^4\:\MS$ on the total mass traced by the MSPs, with the density profile following $\rho_{p}(r) \propto \rho_{\star}(r)^\gamma / \sigma(r),$ with $\gamma = 1.9 \pm 0.3,$  where $\rho_{\star}(r)$ is the stellar mass density and $\sigma(r)$ is the stellar velocity dispersion profile. This favors models in which MSPs form via stellar encounters, such as the leading paradigm whereby MSPs are the progeny of low-mass X-ray binaries.}
{Our analysis demonstrates how combining stellar kinematics with MSP accelerations produces new constraints on mass models, shedding light on the presence or absence of IMBHs at the centers of globular clusters. Further, we provide the first validation of its kind where MSP positions are linked to their place of formation in globular clusters, showing excellent agreement with the expectation of stellar encounter models of MSP formation. This sets a promising precedent amid the advent of the rapidly growing number of observations and discoveries currently being experienced in this field.}  
{}
   \keywords{
  galaxy / globular clusters: individual: Omega Centauri, stars /  pulsars: general, stars: black holes, stars: kinematics and dynamics, galaxies: kinematics and dynamics, galaxies / star clusters: general, cosmology: dark matter}

   \maketitle
%
\section{Introduction}
\label{sec:intro}
The origin and structure of Omega Centauri ($\OC$) is a highly debated topic. Being the most massive and luminous globular cluster (GC) in the Milky Way, it has been argued that $\OC$ may not be strictly speaking a GC, but rather the remnant core of a tidally disrupted, nucleated, dwarf galaxy \citep{Bekki:2003qw, wirth2020formation, johnson2020most}. Evidence in favor of this hypothesis includes the observation of a retrograde orbit \citep{1999AJ....117.1792D}, which may be associated with the disruption of the so-called ``Sequoia'' \citep{myeong2019evidence} or Gaia-Enceladus galaxies \citep{massari2019origin}, the detection of a long tidal stream \citep{ibata2019identification}, and a new formation mechanism for nucleated dwarfs that was recently proposed by \citet{2024arXiv240519286G}. In their model, the smallest nucleated dwarfs have their star formation first quenched by reionization and then reignited in a major-merger driven starburst. The dense nucleus forms in this starburst, yielding a galaxy with two distinct stellar populations that have a large age gap. This observational feature has been seen in several massive Milky Way GCs, including $\OC$ \citep{2024arXiv240519286G}. 


Recent numerical simulations from \cite{wirth2020formation} suggest that, if $\OC$ is indeed an accreted nucleated dwarf, then it can retain significant amounts of dark matter even as it tidally disrupts. By performing a Jeans analysis of stellar kinematics, \cite{Brown:2019whs} and \cite{Evans:2021bsh} found that the presence of an extended dark component in $\OC$ is statistically favored, and that dark matter self-annihilation could account for the observation of $\gamma$-ray signals. However, millisecond pulsars (MSPs) are known $\gamma$-ray emitters and can also account for the observed $\gamma$-ray emission \citep{Reynoso-Cordova:2019biv,Dai:2023pzr, Dai:2019hkk}. 

Alongside the possible presence of dark matter, $\OC$ is interesting as a possible location for the formation of black hole seeds via runaway stellar collisions \citep[e.g.][]{2004Natur.428..724P} or the formation and death of a supermassive star \citep[e.g.][]{2020MNRAS.494.2851C}. Several authors \citep{Noyola:2008kt, 2010ApJ...719L..60N, jalali2012dynamical, 2017MNRAS.464.2174B} found evidence of a $\sim 3 - 5 \times 10^4 \:  \MS$ intermediate-mass black hole (IMBH) at the center of $\OC$ from the consideration of stellar kinematic data (using either Jeans models or N-body simulations). However, \cite{van2010new}, via a Jeans-based analysis, derived a 3$\sigma$ upper bound on the IMBH mass of $1.8 \times 10^4 \; \MS,$ in tension with the previously cited masses. \cite{baumgardt2019no} also found no evidence of an IMBH by comparing N-body simulations to data for $\OC$ (stellar velocity dispersion, surface brightness, and proper motion data), favoring instead a centrally concentrated cluster of stellar-mass black holes that accounts for $4.6 \%$  of the mass of $\OC$. \cite{2017MNRAS.468.4429Z} and \cite{2019MNRAS.482.4713Z} stress that the presence of an IMBH in $\OC$ can be degenerate with orbital anisotropy and with the presence of a cluster of mass-segregated stellar-mass black holes. \cite{Evans:2021bsh} concluded that a core of stellar remnants can account for up to $\sim 5 \times 10^5 \; \MS$ of the inner dark mass derived in their analysis, with a dark matter also being a potential explanation. Therefore, the structure and composition of the mass content of $\OC$ remains a subject of ongoing debate.

In recent years, MSP timing observations have been used to probe the gravitational potential and mass contents of GCs, with a growing wealth of data being produced\footnote{For an updated census of observed MSPs in GCs, we refer the interested reader to this convenient web resource by P.C.C. Freire: \href{https://www3.mpifr-bonn.mpg.de/staff/pfreire/GCpsr.html}{\texttt{https://www3.mpifr-bonn.mpg.de/staff/pfreire/GCpsr.html}}
} and analyzed in this context \citep{ Prager:2016puh,
Freire:2017mgu, 
2017MNRAS.471.1258P,Perera:2017jrk,
2017Natur.542..203K,2017Natur.545..510K, 
Abbate:2018pdf,
2019ApJ...884L...9A,
2020RAA....20..191X,
2021MNRAS.504.1407R, 2021RAA....21..270W,
2023ApJS..269...56Z, Dai:2023pzr, 
2024arXiv240317799P,  2024MNRAS.527.7743X, 2024MNRAS.tmp..838V}. We also refer the interested reader to earlier studies where this methodology was explored, such as in \cite{1993ASPC...50..141P} and \cite{anderson1993study}.

In \cite{Dai:2023pzr}, constraints on accelerations from five centrally located (with projected radii within a few pc from the center) MSPs in $\OC$ were measured  via Doppler-induced effects in timing observations from the pulsars discovered in \cite{Dai:2019hkk}. \cite{Dai:2023pzr} reported a tension between the extremal bounds derived from the baryonic profile used to model the mass content of $\OC$ and the comparatively high inferred pulsar accelerations. This opens the possibility for MSP line-of-sight (LOS) accelerations to play a role in understanding the structure of $\OC,$ adding to the previous discussion. In particular, robust constraints on accelerations near the central region could allow one to discern between the various profiles postulated with unprecedented accuracy, with the ability to identify sufficiently centrally concentrated distributions should they be present. 

With the above in mind, in this work we carry out a Jeans-based analysis exploiting velocity dispersion (proper motion and line-of-sight) and surface brightness profile data with an implementation based on the publicly available code \texttt{GravSphere\footnote{\href{https://github.com/justinread/gravsphere}{\texttt{https://github.com/justinread/gravsphere}}}} \citep{2017MNRAS.471.4541R, 2018MNRAS.481..860R, 2020MNRAS.498..144G, 2021MNRAS.505.5686C}. At the same time, we also include novel constraints on both the pulsar distribution and line-of-sight acceleration data from \cite{Dai:2023pzr} as a self-consistent modification to the original likelihood function used in \texttt{GravSphere}. This allows us to fully exploit both the stellar and MSP timing-derived kinematics.\footnote{We note that during the preparation of this manuscript we were made aware of a recent parallel study that builds on work in \cite{2020MNRAS.491..113H} to also combine stellar kinematics with pulsar accelerations to improve mass models \citep{2024arXiv240706274S}. Their approach is complementary to ours. They also include both stellar kinematic constraints and pulsar accelerations in a similar joint likelihood function, but they implement this within the distribution function modeling code {{\tt limepy}} \citep{2015MNRAS.454..576G}. They then apply their method to the 47 Tuc and Terzan 5 GCs, whereas we focus here on $\OC$.}




An additional novel feature of our approach is that we decompose the inner mass distribution within $\OC$ into a potential IMBH and an additional population tracing the observed distribution of MSPs. This allows us to break degeneracies between the mass in remnants and the possible presence of an IMBH, as well as giving us a robust constraint on the spatial distribution of MSPs. This latter can also be used to shed light on the origin of the gamma-ray emission observed in $\OC$~(\cite{Brown:2019whs,Reynoso-Cordova:2019biv,Dai:2023pzr, Dai:2019hkk}).



This paper is organized as follows. In Sec.~\ref{sec:overview_method}, we describe the mass modeling method that we use in this work and the treatment of the stellar kinematics. In Sec.~\ref{sec:stardata}, we discuss the stellar photometric and kinematic data we use. In Sec.~\ref{sec:constraining}, we discuss the pulsar timing data that we use to determine the pulsar accelerations, while in Sec.~\ref{sec:likedef} we address their implementation in our methodology. In Secs.~\ref{sec:fit} - \ref{sec:discussion}, we present our key results and we discuss them in the context of previous studies in the literature of $\OC$. Finally, in Sec.~\ref{sec:conclusion} we present our conclusions.

\section{Jeans modeling of stellar kinematics}
\label{sec:overview_method}
Our mass modeling method is adapted from the Jeans equations solver \texttt{GravSphere}. We start with a brief overview of how the latest version of \texttt{GravSphere} works. A more detailed discussion can be found in \cite{2017MNRAS.471.4541R,2020MNRAS.498..144G,2021MNRAS.505.5686C} and \cite{Julio:2023oyg}.

In essence, \texttt{GravSphere} is designed to solve the Jeans equations under the assumption of a non-rotating, spherically symmetric stellar system which is in a steady state. This can be used to map velocity dispersion and surface brightness profile of collisionless tracer particles, such as stars in self-gravitating objects like galaxies and GCs, to the total underlying mass distribution.


We start by considering the dynamical equation governing the phase-space distribution function $f(\mathbf{x}, \mathbf{v}, t)$ of a collisionless system of particles, e.g.~stars
, given by the collisionless Boltzmann equation:
\begin{equation}
\label{eq:vlasov}
    \frac{\text{d} f}{\text{d} t} = \frac{\partial f}{\partial t} + \nabla_{\mathbf{x}} f \cdot 
    \mathbf{v} - \nabla_{\mathbf{v}} f \cdot \nabla_{\mathbf{{x}}} \Phi = 0,
\end{equation}
where $\Phi$ is the gravitational potential resulting from  
all mass components in the system, and satisfying the Poisson equation: 
\begin{equation}
\label{eq:poisson}
    \nabla^2_{\mathbf{{x}}} \Phi = 4 \pi G \rho,
\end{equation}
where $\rho$ is the total mass density of the system.

Although the distribution function $f$ is normally not directly accessible via observations, one can multiply Eq.~\eqref{eq:vlasov} by powers of the velocity components and integrate in velocity space to arrive at a dynamic equation  in terms of velocity moments which do relate to observable properties (e.g. see \cite{Binney2008, battaglia2013internal}). In particular, taking the steady state solution $\frac{\partial f}{\partial t} = 0$ for the first moment and multiplying by each velocity component v$_j$, one obtains a set of three well-known equations: The Jeans equations \citep{1922MNRAS..82..122J}, which can be expressed as:
\begin{equation}
\label{eq:jeans}
\sum_{i = 1}^3\frac{\partial \nu_{\star} \braket{\text{v}_i\text{v}_j}}{\partial \text{x}_i}
+ \nu_{\star} \frac{\partial \Phi}{\partial \text{x}_j} = 0,
\end{equation}
where $\nu_{\star} (\mathbf{x}) = \int d^3\text{v} f$ is the stellar (tracer particle) number density and $\braket{\text{v}_i\text{v}_j} = \frac{1}{\nu_{\star}}\int d^3 \text{v} \text{v}_i \text{v}_j f $ correspond to the two-point velocity moments, with the brackets ``$\braket{}$'' generally denoting an integral with the distribution function over velocity space.

For the case of a spherically symmetric distribution, one finds that the only non-trivial equation becomes that involving the radial component, which can be expressed in terms of the radial / tangential velocity dispersions (respectively):
\begin{equation}
\label{eq:disp_def}
\sigma^2_{r/t} = 
\braket{\text{v}^2_{r/t}} - \braket{\text{v}_{r/t}}^2 
\end{equation}
to give the spherical Jeans equation:

\begin{equation}
\label{eq:sph_jeans}
\frac{1}{\nu_{\star}} \frac{\partial \nu_{\star} \sigma^2_r}{\partial r} 
+ \frac{2 \beta \sigma^2_r}{r} = -\frac{G M}{r^2},
\end{equation}
where 
\begin{equation}
\label{eq:beta}
\beta \equiv 1 - \frac{\sigma^2_t}{\sigma^2_r}
\end{equation}
is the so-called velocity anisotropy~\footnote{Note that here we use the 1D tangential velocity dispersion. Some authors use the 2D tangential velocity dispersion, in which case the second term in Eq.~\eqref{eq:beta} should be divided by a factor of two.} and we adopt a Newtonian potential with enclosed mass $M(r)$ within the 3D radial coordinate $r$. It should also be understood that $\nu_{\star}, \beta, \sigma_r, \sigma_t$ are all, in general, functions of $r.$

Eq.~\eqref{eq:sph_jeans} can be solved for the quantity $\nu_{\star} \sigma_r^2$ 
yielding the radial velocity dispersion \citep{2005MNRAS.363..705M,1994MNRAS.270..271V}:
\begin{equation}
\label{eq:tot_disp}
\sigma^2_r (r) = 
    \frac{1}{\nu_{\star}(r) g(r)} \int_r^{\infty} \frac{G M (r') \nu_{\star}(r')}{r'^2} g(r') dr',
\end{equation}
where
\begin{equation}
    \label{eq:g}
    g(r) \equiv \exp \Bigg(2 \int \frac{\beta(r)}{r} dr \Bigg).
\end{equation}
The LOS projection of $\sigma^2_r$, which is a commonly used observable (e.g. \cite{2007ApJ...657L...1S}), is then given by \citep{1982MNRAS.200..361B} 
\begin{equation}
    \label{eq:los_disp}
    \sigma^2_{\rm LOS} (R)
    = \frac{2}{\Sigma_{\star}(R)} \int_R^{\infty} 
    \bigg( 1 -  \frac{R^2}{r^2} \beta(r) \bigg)
    \frac{\nu_{\star}(r) \sigma^2_r(r) r}{\sqrt{r^2 - R^2}} dr,
\end{equation}
where $\Sigma_{*}(R)$ corresponds to the stellar surface density at a projected radius $R$, and one multiplies by the geometrical factors corresponding to the contributions of the radial and tangential components along the line of sight, taking a density-weighted average along the line of sight. 

Having discussed some of the basic elements of a Jeans analysis, we now address an issue which arises due to the fact that (even if the photometric profiles of $\Sigma_{\star}, \nu_{\star}$ are well constrained from surface brightness data) there is a clear degeneracy between the anisotropy $\beta$ and the enclosed mass $M(r)$ in Eqs.~(\ref{eq:tot_disp},\ref{eq:los_disp}). This is known as the $M - \beta$ degeneracy (or also, by extension, the $\rho - \beta$ degeneracy, where $\rho(r)$ is the density) and has been extensively discussed in the literature \citep{1990AJ.....99.1548M,Wilkinson:2001ut,2003MNRAS.343..401L,2009MNRAS.395...76D,2017MNRAS.471.4541R}. This problem is a manifestation of the fact that the Jeans equations alone are not a closed system, meaning that they cannot unambiguously constrain these two quantities without more information being provided. 

Following, for example, \cite{2013MNRAS.436.2598W, van2010new}, in order to resolve the mass-anisotropy degeneracy, one can obtain expressions for the projected proper motions along the projected radius $R$ and the tangential component (respectively), these are given by:
\begin{equation}
    \label{eq:disp_R}
        \sigma^2_{\rm PM, \: R} (R)
    = \frac{2}{\Sigma_{\star}(R)} \int_R^{\infty} 
    \bigg( 1 - \beta(r) +  \frac{R^2}{r^2} \beta(r) \bigg)
    \frac{\nu_{\star}(r) \sigma^2_r(r) r}{\sqrt{r^2 - R^2}} dr
\end{equation}
and
\begin{equation}
    \label{eq:disp_t}
        \sigma^2_{\rm PM, \: t} (R)
    = \frac{2}{\Sigma_{\star}(R)} \int_R^{\infty} 
    \bigg( 1 - \beta(r) \bigg)
    \frac{\nu_{\star}(r) \sigma^2_r(r) r}{\sqrt{r^2 - R^2}} dr.
\end{equation}

Furthermore, following \cite{1990AJ.....99.1548M, 2014MNRAS.441.1584R}, to help eliminate the $\rho - \beta$ degeneracy, one can integrate with higher moments of velocity in the collisionless Boltzmann equation, Eq.~\eqref{eq:vlasov}, to obtain two independent observables known as virial shape parameters (VSPs), which are given by 
 \begin{equation}
 \label{eq:vsp1}
 \text{VSP1} = \frac{2}{5} \int_0^{\infty} G M \nu_{\star} (5 - 2\beta)\sigma_r^2 r dr
 = \int_0^{\infty} \Sigma_{\star} \braket{\text{v}_{\rm LOS}^4} R dR
 \end{equation}
 and
 \begin{equation}
 \label{eq:vsp2}
 \text{VSP2} = \frac{4}{35} \int_0^{\infty} G M \nu_{\star}(7 - 6\beta)\sigma_r^2 r^3 dr
 = \int_0^{\infty} \Sigma_{\star} \braket{\text{v}_{\rm LOS}^4} R^3 dR,
 \end{equation}
 where one can obtain data from observations corresponding to the RHS of Eqs.~\eqref{eq:vsp1} and \eqref{eq:vsp2} that can therefore constrain $\beta.$

For the photometric tracer density profile, in this paper, we introduce the generic $\alpha \beta \gamma$ profile, which has been used in a variety of contexts \citep{1990ApJ...356..359H,1996MNRAS.278..488Z,2014MNRAS.441.2986D, 2020MNRAS.499.2912F, Zentner:2022xux} due to its ability to reproduce a diverse range of distributions analytically. This is given by a double power-law model given as follows:

\begin{equation}
\label{eq:plum}
    \nu_{\star}(r) = \frac{\rho_{c}}{(r/r_{c})^{\gamma} (1 + (r/r_{c})^{\alpha})^{(\beta - \gamma)/\alpha}},
\end{equation}
where we have introduced the three exponent variables $\alpha, \beta, \gamma,$ and the scale radius and density $r_c, \rho_c,$ respectively. In previous versions of \texttt{GravSphere}, a summation of mass components given by Plummer sphere profiles \citep{1911MNRAS..71..460P} was used. Plummer models are commonly used in this context and have been found to provide adequate fits to photometric profiles in many cases, while having a simple analytical form. However, we found the model we used was able to reproduce the observed surface brightness profile significantly better, particularly at larger projected radii where the profile becomes shallower than what Plummer-based models allow.  

The projected tracer surface density, is then given by:
\begin{equation}
\label{eq:surface}
    \Sigma_{\star}(R) = \int_{-\infty}^{\infty}  \nu_{\star}(r(l)) \: dl,
\end{equation}
where, for carrying out the integral, $l$ corresponds to the 
coordinate along the LOS 
so that $r(l) = \sqrt{R^2 + l^2}$. Unlike the original version of {\tt GravSphere}, to aid computational efficiency, we pre-compute this projection before fitting the kinematic and acceleration data, thereby assuming that the uncertainty on the photometric light profile is small as compared to the uncertainty on the velocity dispersion profiles. We explicitly checked that this is a good approximation for $\OC$. 


To obtain the corresponding (cumulative) mass profile $M_{\star}(r)$, one can simply integrate Eq.~\eqref{eq:plum} over space and multiply by a normalization factor needed to match the total mass of the distribution $M_{\star}.$ This factor corresponds to the mass-to-light ratio (if using luminosities rather than number densities or arbitrary units for the surface density profile), which is assumed to be constant throughout the profile.



For the full mass modeling, besides the photometric mass profile $M_{\star}(r)$, which accounts for the dynamical mass of the photometric distribution, including stars and any remnants or objects traced by this distribution, we add a central mass, $M_{\rm cen}(r)$, which emulates a generic central cluster of remnants. This central mass is modeled as a single Plummer sphere component \citep{1911MNRAS..71..460P} given by 
\begin{equation}
\label{eq:masscen}
    M_{\rm cen}(r) = M_{\rm cen} \frac{r^3 }{r^3_{\rm cen}} \Bigg( 1 + \frac{r^2}{r^2_{\rm cen}}
    \Bigg)^{-3/2},
\end{equation}
where in this case we introduce the corresponding total mass and scale lengths $M_{\rm cen}$ and $\: r_{\rm cen}$, respectively. To model the presence of an IMBH, we also introduce a point mass $M_{\rm BH}.$

Lastly, since the full likelihood implementation of MSP data requires a model of the MSP distribution (see  Sec.~\ref{sec:likelihood}), we also include a mass component that follows the MSP profile. This allows us to consider the presence of a distribution of stellar remnants traced by the MSPs that are more centrally concentrated than the stars. This profile is given by
\begin{multline}
\label{eq:massmsp}
    M_p(r) = 
    \frac{4 M_p}{3 \sqrt{\pi}} 
    \frac{\Gamma \Big[ (1 - \alpha)/2\Big]}
    {\Gamma \Big[ -(\alpha/2 + 1)\Big]}
    \frac{r^3}{r_0^3} \:
    {_2F_1} \Bigg[ \frac{3}{2}, \: \frac{1 - \alpha}{2}; \: \frac{5}{2}; \:
    -\frac{r^2}{r_0^2} \Bigg],
\end{multline}
where ${_2F_1}$ is the Gaussian hypergeometric function and $M_p$ is the total mass of the distribution, which is defined (convergent) for the density exponent parameter $\alpha < -2$, while $r_0$ is the length scale parameter. This equation is obtained by integrating the density with the functional form shown in Eq.~\eqref{eq:pul_3d} over space and multiplying it by the normalization factor to obtain the mass 
distribution. 
The resulting total mass profile is thus given by:
\begin{equation}
\label{eq:massmod}
    M(r) = M_{\star}(r)  +  M_{\rm cen}(r) + M_{\rm BH} + M_p(r).
\end{equation}
While dark remnant models of $\OC$ favor two-component profiles of light and heavy remnants (see discussion in Sec.~\ref{sec:rem} and references therein), in general, different populations of objects are expected to show different degrees of segregation based on their dynamical histories and their intrinsic masses. Being more massive than main-sequence stars, but less than heavy remnants (such as stellar-mass black holes), MSPs could indeed trace a distribution of intermediate concentration, should the system undergo sufficient mass segregation (e.g. \cite{1993ASPC...50..141P}).  Here we stress the role of pulsars is to act as tracers, rather than direct contributors to the kinematics. This mechanism is allowable, for instance, if other remnants of similar mass which are likely more abundant, such as white dwarfs and neutron stars, would undergo a similar level of mass segregation and thus be traced by the pulsars, even if these have negligible kinematic contributions by themselves. On the other hand, if more limited mass segregation has taken place, then these other remnants could be traced by the photometric profile instead (showing little segregation as with the lighter main-sequence stars) and only the heavy remnants (e.g. stellar-mass black holes) could segregate (as predicted by some models and simulations, see Sec.~\ref{sec:rem}). In this case, a more concentrated distribution of pulsars could be explained by their mechanism of formation rather than mass segregation (see Sec.~\ref{sec:mspan}).

There is also no a priori reason to exclude the coexistence of an IMBH with a central remnant distribution, which is an important consideration given that both have been invoked to explain $\OC$'s kinematics. 
Therefore, to fully explore all these scenarios, their potential kinematic relevance, and to explicitly consider any potential degeneracies between them, we have decided to work with the generic multi-component model from Eq.~\eqref{eq:massmod}.\footnote{We note also that our choice of mass modeling is very general in that it makes few assumptions about the underlying distribution by allowing for multiple different mass components which are varied in nature and morphology. We also performed an additional analysis with different dark component parametrizations (Appendix~\ref{app:diffdark}), finding that our main results are not affected by these.}

We use the anisotropy profile assumed in \texttt{GravSphere} following the generalized form from \cite{2007A&A...471..419B}, which is an extension of the Osipkov-Merritt anisotropy profile \citep{1979PAZh....5...77O,1985MNRAS.214P..25M}, and is given by
\begin{equation}
    \label{eq:beta_ansatz}
    \beta(r)  = \beta_{0} 
    + (\beta_{\infty} - \beta_0) \frac{1}{1 + \Big(\frac{r_t}{r}\Big)^{\eta}},
\end{equation}
where $\beta_0$ is the anisotropy at the center, $\beta_{\infty}$ its limit as it asymptotically approaches infinity, $r_t$ is the transition scale between these two regimes, and $\eta$ is the exponent which modulates the steepness of the transition. 

\section{Stellar kinematic and photometric data}\label{sec:stardata}

We combine the LOS stellar kinematic data from \citet{2006A&A...445..503R} with \citet{2009MNRAS.396.2183S}, adjusting their individual line of sight velocities to match the value in \citet{2024MNRAS.528.4941P}, $\langle v_{\rm LOS} \rangle = 232.5$\,km/s. This accounts for any systematic offset between the different datasets (the shift is of order 1\,km/s, which is smaller than the uncertainty on the individual stellar velocities). We perform a quality cut on these data, retaining only those stars with velocity error smaller than 4\,km/s. This yields 1644 LOS velocities. We augment these with the \citet{2010ApJ...719L..60N} data that are pre-binned for three different choices of center; more on this, below. We take HST proper motion (PM) data from \citet{2017ApJ...842....6B}, augmenting this with Gaia proper motion data from \citet{2021MNRAS.505.5978V}. As with the LOS data, we perform a similar quality cut on the PM data, retaining stars with PM uncertainties smaller than 2\,km/s at the distance of $\OC$ (assumed to be 5.2\,kpc; see Sec.~\ref{sec:constraining}), this leaves a combined total of $\sim 200,000$ PMs which pass this criterion (including also additional filters based on data quality and membership probabilities). The PM data are perspective corrected as in \citet{2006A&A...445..513V}, assuming a mean LOS velocity and distance, as above. 
The surface brightness data are taken from the \citet{2017ApJ...842....6B} catalog and augmented with Gaia photometry from \cite{2019MNRAS.485.4906D}.
We explore three different choices of center for $\OC$ taken from \citet{anderson2010new} (${\rm RA, DEC} = 201.69683333,  -47.47956944$), \citet{Noyola:2008kt}  (${\rm RA, DEC} = 201.69184583,  -47.47911111$) and the kinematic center from \citet{2010ApJ...719L..60N} (${\rm RA, DEC} = 201.69630208, -47.47835389$). These centers are self-consistently computed for the photometric light profile, LOS and PM stellar kinematic data.

The LOS velocity and PM data are then binned using the {\tt binulator}, as described in \citet{2021MNRAS.505.5686C}. We use 100 logarithmically spaced bins, ensuring that no bin has less than 100 stars in it. For the LOS data, we augment the post-binned data with the pre-binned data from \citet{2010ApJ...719L..60N}, using the consistent center reported in that work. {\tt binulator} calculates the two virial shape parameters, VSP1 and VSP2 (see Sec.~\ref{sec:meth}) and their uncertainties. As such, we use these also in our fits to assist with breaking the mass-anisotropy degeneracy (see Sec.~\ref{sec:overview_method}). We explicitly checked whether our choices of binning parameters impact our results. Replacing our binned PM data with the binned data from \citet{2015ApJ...803...29W} yielded similar results for our mass components, with the main difference being a somewhat more extended distribution for the central mass component, which had a comparable total mass, showing no significant differences for the IMBH component. The stellar kinematic data and the best-fit photometric parameters used in this study have been made publicly available.\footnote{\href{https://github.com/dadams42/OCenKinematics}{\texttt{https://github.com/dadams42/OCenKinematics}}}


\section{Pulsar timing observations}
\label{sec:constraining}
Although pulsars are known to be remarkably regular sources, over sufficiently long time scales, it is in fact possible to detect significant variations in the observed period of these objects. This effect encodes information on the relative acceleration between the observer and pulsar reference frames, as well as the intrinsic spin-down of the pulsar (e.g. see \cite{1993ASPC...50..141P}), and can be calculated through the following equation:
\begin{equation}
\label{eq:pobs}
\Bigg(\frac{\dot{P}}{P}\Bigg)_{\rm obs} = \frac{a_{\rm LOS} + a_{\rm S}+ a_{\rm g} }{c} + \Bigg(\frac{\dot{P}}{P}\Bigg)_{\rm int}, 
\end{equation}
where $c$ is the speed of light and $P_{\rm obs/int}, \: \dot{P}_{\rm obs/int}$ denote the period and its time derivative in the observer / pulsar rest frames (respectively), with the second term in the RHS corresponding to the intrinsic spin-down contribution of the pulsar. The first term in the RHS accounts for the various contributions due to the relative motions between the pulsar and the observer which induce an effective time-dependent Doppler shift in the observed period. We will now address the various components which constitute this term. 

$a_g$ is the relative difference in LOS accelerations due to the differential Galactic rotations of the GC and the Solar System. Following \cite{Dai:2023pzr, Freire:2017mgu} and Section 5.1.2 of \cite{Prager:2016puh}, we use Equation 5 from \cite{1995ApJ...441..429N} yielding 
\begin{equation}
\label{eq:a_g1}
a_g = - \cos(b) \: \Bigg(\frac{\Theta_0^2}{R_0} \Bigg) \: \Bigg( 
\cos(\ell) + \frac{\vartheta}{\sin^2(\ell) + \vartheta^2} \Bigg) \rm \; \; m \, s^{-2},
\end{equation}
where $\vartheta \equiv (d/R_0)\cos(b) - \cos(\ell).$ $R_0 = 8.34 \pm 0.16$ kpc is the distance from the Sun to the Galactic center and $\Theta_0 = 240 \pm 8$ km/s is the Galactic rotational speed at that point, both of which are obtained from \cite{2014ApJ...783..130R}. The distance from the cluster is fixed to $d = 5.2$\,kpc. This value was originally chosen for consistency with the values reported in \cite{2014ApJ...797..115B, 2015ApJ...803...29W, 2015ApJ...812..149W}, and was subsequently validated throughout our analysis (see Sec.~\ref{sec:meth}). We use the Galactic coordinates $\ell = 309.1  ^\circ, $ $ b = 14.97 ^\circ$ corresponding to the center of $\OC$ located at $\rm RA = $~13:26:47.24, DEC = -47:28:46.5 adopted in \cite{Dai:2023pzr}.\footnote{This center was originally reported in \cite{1996AJ....112.1487H} catalog (2010 edition). It is essentially equivalent to the \citet{anderson2010new} center, which is the main center we adopt and present in this analysis for the stellar kinematic and photometric data. For the remaining centers (Appendix~\ref{app:centers}), given the relatively extended positions of the MSPs and that these are not dependent on binning prescriptions, the choice of center should have a limited effect on the results obtained.}

For the propagation of errors, we used the $68 \%$ confidence limit (CL) interval about the median obtained after running one million iterations, where a set of random samples is generated for the input parameters $\Theta_0, R_0,$ under the assumption that they are independent and normally distributed with median values and errors as cited.

$a_S$ accounts for the proper motion of the GC which induces an apparent LOS acceleration contribution -- the Shklovskii effect -- which is given by \citep{shklovskii1970possible, Dai:2023pzr}
\begin{equation}
\label{eq:a_S1}
    a_S = 3.78 \times 10^{-12} \: \Bigg( \frac{d}{5.2 \; \rm kpc} \Bigg) \:
    \Bigg( \frac{\mu_T}{\rm mas \: yr^{-1}} \Bigg)^2 \; \rm m \, s^{-2},
\end{equation}
where $\mu_T = \sqrt{\mu_{\delta}^2 + \mu_{\alpha *}^2}$ is the proper motion, and we take
$\mu_{\delta} = -6.7445 \pm 0.0019 \: \rm mas \, yr^{-1}, $  $ \mu_{\alpha *} = -3.1925 \pm 0.0022 \: \rm mas \, yr^{-1} $ \citep{helmi2018gaia}. 

$a_{\rm LOS}$ corresponds to the LOS component of the acceleration caused by the gravitational field of the GC on the pulsar and, under the assumption of spherical symmetry, is given by
\begin{equation}
\label{eq:a_gc}
a_{\rm LOS} (r, l) = - G \, \Bigg( \frac{l}{r} \Bigg) \, 
\frac{M(r)}{r^2},
\end{equation}
where $l$ in this case is the longitudinal component (i.e. along the line of sight) of the position of the pulsar from the center of the GC and $r$ is the total distance, so that $r = \sqrt{l^2 + R^2},$ where $R$ is the projection orthogonal to the line of sight and is typically the one that can be measured directly. $M(r)$ denotes the enclosed mass of the spherical density distribution, while the geometric factor $l/r$ projects the LOS component of the total radial acceleration experienced by the pulsar. This is the component of the observed period derivative which traces the mass contents of the GC, and is therefore the reason why these timing observations are of interest for the present study. 

While the intrinsic spin-down component is subject to significant uncertainties and is difficult to measure directly, it can, however, be approximated to a model of magnetic dipole emission with breaking index $n = 3,$ yielding \citep{Prager:2016puh, Dai:2023pzr}
\begin{equation}
\label{eq:dipole}
    a_{\rm int} \equiv c \, \Bigg( \frac{\dot{P}}{P} \Bigg)_{\rm int} \approx 
    7.96 \times 10^{-10} \, \Bigg( \frac{B}{2 \times 10^8 \rm \: G} \Bigg)^2 \,
    \Bigg ( \frac{2 \rm ms}{P_{\rm obs}} \Bigg)^2 \: \rm m \, s^{-2},
\end{equation}
where $B$ is the surface magnetic field strength of the pulsar, whose effective values can be constrained from observations where this term is dominant (see Sec.~\ref{sec:like_priors}).

\section{Likelihood definition}
\label{sec:likedef}

\subsection{Implementation of stellar kinematics constraints}

For the purposes of fitting the stellar kinematics observables discussed in Sec.~\ref{sec:overview_method}, one can construct a general log-likelihood function for the proper motion and LOS velocity dispersions, and the virial shape parameters, taking the form:
\begin{equation}
    \label{eq:llike_gen}
    \ln \mathcal{L}\big(\boldsymbol{\theta} \big) = -\frac{1}{2} \sum_y \chi_y^2
    ,
\end{equation}
where we have the general chi-square variable for an observable $y$ defined as
\begin{equation}
    \label{eq:chi2}
    \chi_{y}^2 \equiv \sum_i \frac{\bigg[y_{i \rm, \: obs} - y_{i} \big(\boldsymbol{\theta} \big) \bigg]^2}{\delta y_i^2},
\end{equation}
and observables are taken to be normally distributed with median values $\{y_{i \rm,  \: obs} \}$ and corresponding standard errors $\{\delta y_i\},$ and $\{y_{i} \big(\boldsymbol{\theta} \big)\}$ correspond to the predicted variables as a function of the mass and (symmetrized) anisotropy model parameters $\boldsymbol{\theta}.$

\subsection{Implementation of pulsar constraints}
\label{sec:likelihood}

There are various possible ways to implement the pulsar timing constraints. However, there are some subtleties that should be considered when doing so. For example, a pointwise optimization for each of the pulsars over the nuisance parameter $l$ can be quite numerically expensive if this has to be done for each step of a fitting routine. Also, errors or confidence intervals in acceleration constraints are not originally given in \cite{Dai:2023pzr} and can be difficult to quantify, particularly when including the intrinsic spin-down contribution.

To allow for an effective implementation and with the above discussion in mind, we include in the fit the quantity:
\begin{equation}
\label{eq:aobs}
    a_{\rm obs} \equiv  \, c \: \Bigg(\frac{\dot{P}}{P}\Bigg)_{\rm obs}  - a_g - a_S 
    = a_{\rm LOS} + c\: \Bigg(\frac{\dot{P}}{P}\Bigg)_{\rm int} ,
\end{equation}
where the RHS follows directly from Eq.~\eqref{eq:pobs} and corresponds to the theoretical prediction from the GC mass model used, including the contribution from the intrinsic spin-down in Eq.~\eqref{eq:dipole}. This observable is well constrained in each of its components. The errors can be propagated following our approach in Sec.~\ref{sec:constraining}: We simulate one million samples assuming that input variables are independent and normally distributed, with the specified errors and median values taken from the literature (see Sec.\ref{sec:constraining}). We checked explicitly that the resulting distributions of $a_{\rm obs}$ were close to normal and obtained similar values to those predicted by common standard error propagation techniques (i.e. applying the central limit theorem to linear expansions of the functions). The resulting median values and errors on $a_{\rm obs},$ as well as some of the data of the MSPs published in \cite{Dai:2023pzr},
are shown in Table~\ref{tab:p_errors}.

\begin{table*}[t]
\caption{Data from MSPs published in \cite{Dai:2023pzr} with the inclusion of our derived median values and  $68 \%$ CL errors of $a_{\rm obs}$, as defined in Eq.~\eqref{eq:aobs}, as well as the central values of $R$ (see discussion following Eq.~\eqref{eq:a_gc} for details).} 
    \centering
    \setlength{\tabcolsep}{0.7em}
\renewcommand{\arraystretch}{1.5}
  \setlength{\arrayrulewidth}{.30mm}
    \begin{tabular}{c|ccccccc}
     \hline\hline
  Name & $P_{\rm obs}$  & $\dot{P}_{\rm obs}$  & RA & DEC & $R$ & $a_{\rm obs}$ \\
  &[ms]&[$10^{-20}$ s\,s$^{-1}$]&[J2000]&[J2000]&[pc]&[$10^{-9}$ m$\,$s$^{-2}$]\\ \hline
  J1326-4728A &  4.108786192190(1) & 2.738(2) &  13:26:39.6699(2) & -47:30:11.641(3) & 2.89 & $1.914 \pm 0.011$\\
  J1326-4728B &  4.791869161014(2) & -5.433(4) &  13:26:49.5688(3) & -47:29:24.889(4) & 1.14  & $-3.483 \pm 0.011 $\\
  J1326-4728C &   6.867859692327(6) & 0.98(1) &  13:26:55.2219(6) & -47:30:11.753(9) & 2.96 & $0.344 \pm 0.012$\\
 J1326-4728D &   4.578833468410(2) & -4.110(4) &  13:26:32.7130(2) & -47:28:40.053(3) & 3.72 & $-2.775 \pm 0.012$\\
 J1326-4728E &  4.2077170407405(4) & 1.628(1) &  13:26:42.67844(7) & -47:27:23.999(1) & 2.38 & $1.076 \pm 0.011$\\    
       \hline
         \hline
         \end{tabular}
         \tablefoot{Error propagation follows the approach outlined in Sec.~\ref{sec:constraining}.}
\label{tab:p_errors}
         \end{table*}

The term to be added to the log-likelihood function takes the form:
\begin{align}
\label{eq:llhood}
       \ln \mathcal{L}_{p, \: \rm LOS} (\boldsymbol{\theta}_M; \, \{ l_i \}, \{B_i \}) = -\frac{1}{2} \sum_i^{N_p} \frac{(a_{\text{obs}, i } - a(\boldsymbol{\theta}_M; \, l_i, B_i))^2 }{\delta a_i^2 }, 
\end{align}
where $a(\boldsymbol{\theta}_M; \, l_i, B_i)$ is the predicted LOS acceleration including intrinsic spin-down (RHS of Eq.~\eqref{eq:aobs}) and $\boldsymbol{\theta}_M$ are the set of parameters characterizing the mass model. The parameters $\{B_i \}$ and $\{l_i\}$ correspond to the (unknown) magnetic field and longitudinal components of each of the  $N_p$ pulsars (5 in our case) while $\{ \delta a_i \}$ are the propagated errors of Table~\ref{tab:p_errors}.

To perform the fit, we marginalize over the parameters $\{B_i \}$ and $\{l_i\}$. 
The $l_i$ are allowed to vary over a wide range within the boundaries of the GC. The reason $B$ is marginalized over is that it has not been measured for this sample of MSPs. However, a reasonable idea of its range can be obtained based on known measurements performed on similar MSP populations. This is possible by measuring period variations for MSPs with no GC potential contributions, meaning that, after correcting for differential Galactic rotation and the Shklovskii effect, these observations act as direct tracers of the intrinsic spin-down contributions. For binary systems that do not experience significant orbital variability, Doppler-induced changes can also be studied in orbital period derivatives, with the added benefit of being independent of intrinsic spin-down effects (e.g. \cite{Freire:2017mgu,Prager:2016puh}). To date, of the 18 known MSPs discovered in $\OC$, only the 5 used in this study have measured spin period derivatives and the only known binary from this set has not had its orbital period derivative measured \citep{Dai:2023pzr}. Therefore, we resort to existing studies of comparable MSP populations to constrain the intrinsic spin-down components. In particular, \cite{Prager:2016puh} found using data from spin-dominated MSP populations in the ATNF catalog \citep{2005yCat.7245....0M} that the inferred distribution of $B$ was well approximated to be log-normal with median $\log_{10} B \: [ {\rm G} ] = 8.47$ and standard deviation of 0.33\,dex, which we adopt as a prior.
This allows for an ample range which captures some of the uncertainty in the intrinsic spin-down, while also being representative of the values expected for MSP populations.

$a_{\rm obs}$ can be regarded as a strict upper bound on the acceleration, in the sense that the contribution due to intrinsic spin-down will always be negative. This translates to a lower bound on the enclosed mass when the accelerations are negative. Implementing this as a prior during the fits would be a somewhat simpler approach and has the advantage of not depending on the uncertainties in the intrinsic spin-down. However, it is likely less constraining and would require the potentially numerically expensive pointwise optimization over the nuisance parameters $\{l_i\}$ to determine whether the bound is satisfied.

To fully exploit the MSP data and help further constrain parameters, we adopt an approach similar to the one used by \cite{Prager:2016puh} and \cite{Abbate:2018pdf}. We add a positional component to the likelihood function, adopting the ``generalized King model'' profile which is typically used to model MSP populations. This has a surface density given by \citep{1995ApJ...439..191L}:
\begin{equation}
    \label{eq:pul_surf}
    n(R) = n_0 \Bigg[1 + \bigg(\frac{R}{r_0}\bigg)^2 \Bigg]^{\alpha / 2},
\end{equation}
where we have two additional free parameters: the exponent $\alpha,$ which modulates the steepness of the distribution; and the scale radius $r_0$. (We also have the additional parameter $n_0,$ which is the central surface density. However, this is not included in the fit as it is a global factor and the likelihood components that we shall introduce are independent of it.) 

As seen in \cite{Grindlay:2002fb}, the 3D density has the form:
\begin{equation}
    \label{eq:pul_3d}
    n(r) = 
    f(\alpha, \: r_0, \: n_0)
    \Bigg[ 1 + \bigg(\frac{r}{r_0} \bigg)^2 \Bigg]^{(\alpha - 1)/2},
\end{equation}
where $f(\alpha, \: r_0, \: n_0)$ is a normalization factor that can be determined using the relation of the projected surface density (as in Eq.~\eqref{eq:surface}) so that 
\begin{equation}
\label{eq:proj_pul}
   n(R) = \int_{-\infty}^{\infty} n\Bigg(\sqrt{R^2 + l^2} \Bigg) \: dl,
\end{equation}
and one can verify that this expression yields Eq.~\eqref{eq:pul_surf} with 
\begin{align}
    \label{eq:falpha}
    f(\alpha, \: r_0, \: n_0)= 
\frac{\Gamma[ (1 - \alpha) /2 ]}{\sqrt{\pi} \Gamma(-\alpha/2)} \frac{n_0}{r_0},
\end{align}
where $\Gamma$ denotes the Euler gamma function.

To quantify the positional probability to be added to the likelihood function, we follow a similar approach to Appendix D of \cite{anderson1993study} and \cite{1993ASPC...50..141P,Prager:2016puh, Abbate:2018pdf}. We express the total positional probability density (for an individual pulsar) as:
\begin{align}
\label{eq:pos_prob}
    P(l, R | \alpha, r_0) =  P(l\,| R , \alpha, r_0) P(R|  \alpha, r_0) ,
\end{align}
where: 
\begin{align}    
\label{eq:conprob_l}
    P(l\,| R, \alpha, r_0) &\: dl
    =  \frac{ n\Bigg( \sqrt{R^2 + l^2} \Bigg) \: dl}
    {\int_{-\infty}^{\infty}  n\Bigg(\sqrt{R^2 + l^{\prime2}} \Bigg) \: dl^{\prime}}
    \notag\\
   & = \frac{\Gamma[ (1 - \alpha) /2 ]}{\sqrt{\pi} \Gamma(-\alpha/2)} \frac{1}{r_0} \frac{\Big[ 1 + \Big(R^2 + l^2\Big)/r_0^2 \Big]^{(\alpha - 1)/2} }{\Big[1 + (R/r_0)^2 \Big]^{\alpha / 2}} \: dl,
\end{align}
and:\footnote{Note that, in our definition, $R$ is  non-negative, so that we only integrate from 0 outwards (extending this to negative values would only change the integral by a global factor of 2, which would not affect our results).}
\begin{align}    \label{eq:conprob_rp}
    P(R| \alpha, r_0)\: dR&=  \frac{R n( R)\: dR}
    {\int_{0}^{\infty}  R' n(R' ) \: dR'} \notag\\
    &= - (\alpha + 2) \frac{R}{r_0^2} \Bigg[1 + \bigg(\frac{R}{r_0}\bigg)^2 \Bigg]^{\alpha / 2}\: dR,
\end{align}
where in Eq.~\eqref{eq:conprob_l} we divide the infinitesimal amount of pulsars expected in a 3D ring of radius $R,$ width $dR,$ and length $dl$ by those in the infinite cylinder of equal radius and width, recovering Eq.~(3.7) of \cite{1993ASPC...50..141P}. Similarly, for Eq.~\eqref{eq:conprob_rp} we do the 2D analog, with the surface element of an annulus of width $dR$ and radius $R,$ integrating the whole disk with the projected surface density to obtain the relevant normalization factor in the denominator.

The complete positional probability for a given pulsar is thus given by:
\begin{align} \label{eq:comb_prob}
P(l, R | \alpha, r_0)= -\frac{\alpha + 2}{\sqrt{\pi}}
\frac{\Gamma \Big[ (1-\alpha)/2\Big]}{\Gamma \Big[-\alpha/2\Big]}     \frac{R}{r_0^3}
     \bigg[1 + (l^2 + R^2)/r_0^2 \bigg]^{(\alpha - 1)/2},
\end{align}
yielding the relevant (i.e., non-constant) contribution to the log-likelihood function for $N_p$ pulsars:
\begin{align}
    \label{eq:llike_pos}
   \ln \mathcal{L}_{\rm pos}&(\{ l_i \}, \alpha , r_0)
   = \frac{\alpha - 1}{2} \sum_i ^{N_p}
   \ln (1 + (R_i^2 + l_i^2)/r_0^2) \nonumber\\
   &+ N_p \Bigg( \ln \frac{\Gamma \Big[ (1-\alpha)/2\Big]}{\Gamma \Big[-\alpha/2\Big]} + \ln[-(\alpha + 2)] - 3 \ln r_0 \Bigg).
\end{align}
Since \cite{Chen:2023lzp} have also reported the positions of 13 additional MSPs, but without the relevant timing data that trace kinematics, one can still use these positions to constrain the MSP distribution without the need to introduce any additional free parameters. For this purpose, we focus only on the projected radius component of the likelihood (as  $l$ becomes a redundant parameter). This gives the following log-likelihood (for $N_p'$ pulsars):
\begin{align}
\label{eq:ll_perponly}
\ln \mathcal{L}_{\rm pos, \: R }(\alpha , r_0)&= \frac{\alpha}{2} \sum_i^{N_p'} \ln \Bigg[1 + \bigg(\frac{R_i}{r_0}\bigg)^2 \Bigg] \nonumber\\
&+
 N_p' ( \ln [-(\alpha + 2)] - 2 \ln r_0).
\end{align}
Note that in our treatment of the likelihood implementation of these MSP observables we have been careful to include normalization factors that depend on the parameters $\alpha$ and $r_0.$ While these are not always explicitly shown in previous works (and can sometimes be ignored), they are necessary to include whenever they are parameter-dependent and not merely global factors of the likelihood or, correspondingly, additive constants in the log-likelihood function. 


\begin{table}[t]
\caption{Priors implemented in \texttt{dynesty} for our analysis.} 
    \centering
    \setlength{\tabcolsep}{0.7em}
\renewcommand{\arraystretch}{1.5}
  \setlength{\arrayrulewidth}{.30mm}
    \begin{tabular}{c|c c c}
     \hline\hline
        \textbf{Parameter} & \textbf{Units} & \textbf{Prior Type} & \textbf{Range}  \\ \hline 
        $\widetilde{\beta}_0$ & none & flat &  $[-1, \; 1]$
        \\
        $\widetilde{\beta}_{\infty}$ & none & flat &  $[-1, \; 1]$
        \\
        $\log_{10} r_t$ & pc & flat &  $[0, \; 2]$
        \\
        $\eta$ & none & flat &  $[0, \; 4]$
        \\
        $r_{\rm cen}$ & pc & flat &  $[0.5,  \; 2.5]$
        \\
        $\log_{10} M_{\rm cen}$ & $\MS$ & flat &  $[0,  \; 6]$
        \\
        $\log_{10} M_{\rm BH}$ & $\MS$ & flat &  $[0,  \; 6]$
        \\
        $\log_{10} M_{\star}$ & $\MS$ & flat &  $[5, \;  8]$
        \\
        $\log_{10} M_{p}$ & $\MS$ & flat &  $[0,  \; 6]$
        \\
        $\alpha$ & none & flat &  $[-8,  \; -2]$
        \\
        $r_0$ & pc & flat &  $[0,  \; 10]$
        \\
        $\log_{10} B_i$ & G & normal & 
       \begin{tabular}{@{}c@{}}$\mu = 8.47 $ 
       \\  $  \sigma = 0.33$ 
       \end{tabular}
       \\
       $l$ & 4.54 pc & flat &  $[-10,  \; 10]$
        \\
        \hline\hline
    \end{tabular}
\tablefoot{To ensure a broad range of values is considered, the variable $\log_{10} B_i$~[G] is sampled over the flat range [6, 10], inside which the normal prior is applied. See discussion in text for more details on the priors adopted. }
\label{tab:priors}
\end{table} 

\subsection{Full likelihood and priors}
\label{sec:like_priors}

The full log-likelihood can now be summarized as 
\begin{align}
    \label{eq:full_ll}
    \ln \mathcal{L}_{\rm tot} (\boldsymbol{\theta}) &=
    \ln \mathcal{L}_{\rm LOS} +
       \ln \mathcal{L}_ {\rm PM, t} + \ln \mathcal{L}_{\rm PM, R}  +
        \ln \mathcal{L}_{\rm VSP1} 
        \notag\\
        &+  \ln \mathcal{L}_{\rm VSP2}     
       + \ln \mathcal{L}_{p, \: \rm LOS}
       + \ln \mathcal{L}_{\rm pos} + \ln \mathcal{L}_{\rm pos, \: R }, 
\end{align}
where $\boldsymbol{\theta}$ includes all the model parameters which are simultaneously fitted, including anisotropy, photometric profile, MSP distribution, and mass model parameters. This is an important aspect of this study, as it allows for a fully consistent statistical treatment of the data and exploring potential degeneracies and correlations between different parameters, such as the aforementioned mass-anisotropy degeneracy. This interdependence of parameters may not otherwise be evident if one were to make separate fits or fix some of these parameters implicitly. Importantly, as will be shown in subsequent sections, it also allows for a full exploitation of the constraints imposed by the data in a way that differs from the results obtained from a separate treatment.


The prior on $l$ is expressed in terms of the core-radius length scale $r_c \equiv 4.54$~pc of \cite{baumgardt2018catalogue}, which is also used by \cite{Dai:2023pzr}. Following \cite{2017MNRAS.471.4541R}, we also work with the symmetrized anisotropy:
\begin{equation}
\label{eq:beta_t}
\widetilde{\beta} \equiv \frac{\beta}{2 - \beta} 
\end{equation}
for $-1\leq \widetilde{\beta} \leq 1$ to efficiently sample the (infinitely) broad range of possible values of  $\beta.$


Table~\ref{tab:priors} shows the priors we use in our analysis, whose results will be discussed in the subsequent sections. The prior on $\alpha$ is based on the range of physically expected limits for pulsar populations, following the discussion in \cite{1993ASPC...50..141P, Prager:2016puh}.\footnote{We found that extending these limits within the physically allowable range that allows for convergence and positivity of the distribution does not greatly affect its overall shape due to the degeneracy in the parameters $r_0$ and $\alpha.$} These limits are chosen to be broad and away from the tails of the posterior distribution in most cases, so that results are insensitive to them, while being sufficiently narrow to allow for satisfactory numerical efficiency. Priors in log-space are designed to efficiently span a broad range of several orders of magnitude.

The fits are performed using the nested sampling package \texttt{dynesty} \citep{2020MNRAS.493.3132S, 2022zndo...6609296K}. This is based on nested sampling techniques \citep{2004AIPC..735..395S,10.1214/06-BA127} and the subsequently developed dynamic nested sampling technique \citep{cite-key}, which we use, and is optimized for the inference of posterior distributions (as opposed to evidence estimation), using the bounding method from \cite{2009MNRAS.398.1601F}. The sampling method used is based on \cite{10.1214/aos/1056562461,2015MNRAS.450L..61H,2015MNRAS.453.4384H}. These methods are designed for more efficient sampling of complex posterior distributions which may present multimodalities and a potentially large number of dimensions in parameter space. The reason for this choice is that, in these settings, traditional Markov-chain Monte Carlo techniques
tend to have less satisfactory performance, with numerical convergence being harder to achieve and the process being costly in CPU time. Another advantage of \texttt{dynesty} is that it is comparatively simple to implement, with the inclusion of an automated initialization routine 
and stopping criteria.\footnote{For further information, see: \\ \href{https://dynesty.readthedocs.io/en/latest/index.html}{\texttt{https://dynesty.readthedocs.io/en/latest/index.html}}.}


\section{Kinematic fits and constraints on a central mass}
\label{sec:fit}

\begin{figure*}[h!]
    \centering

\includegraphics[width=6.45cm]{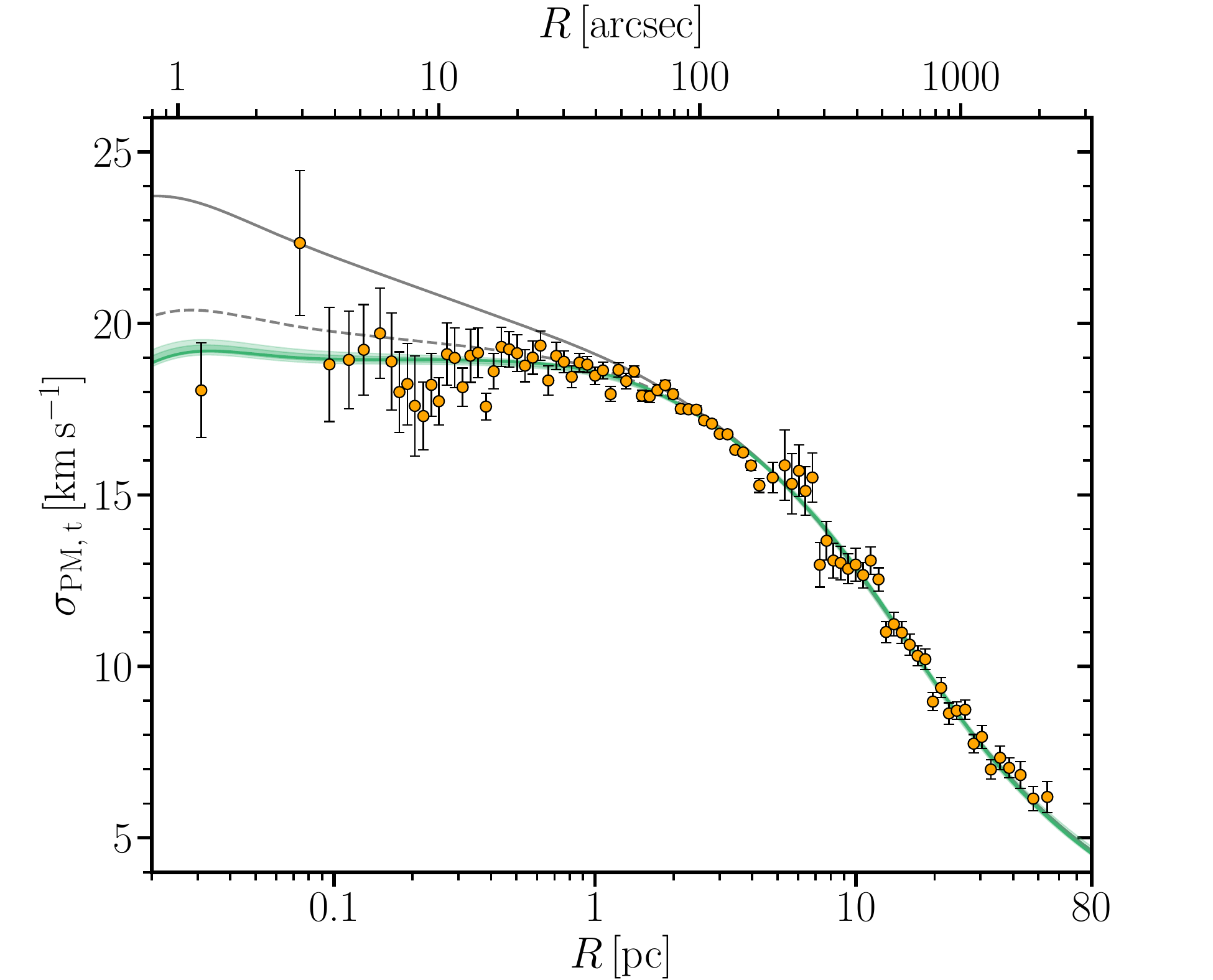} 
\hspace{-0.7cm}
\includegraphics[width=6.45cm]{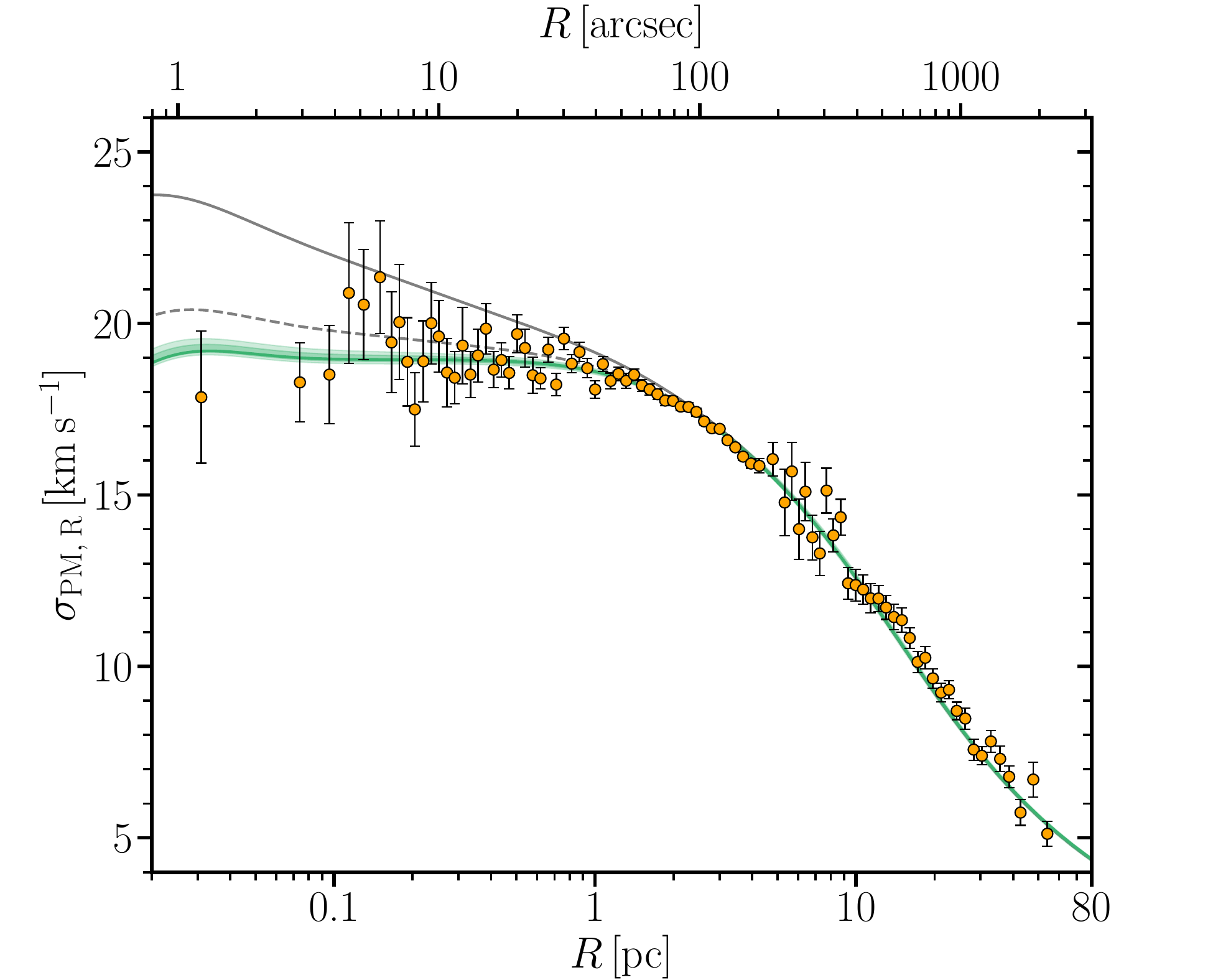}
\hspace{-0.7cm}
\includegraphics[width=6.45cm]{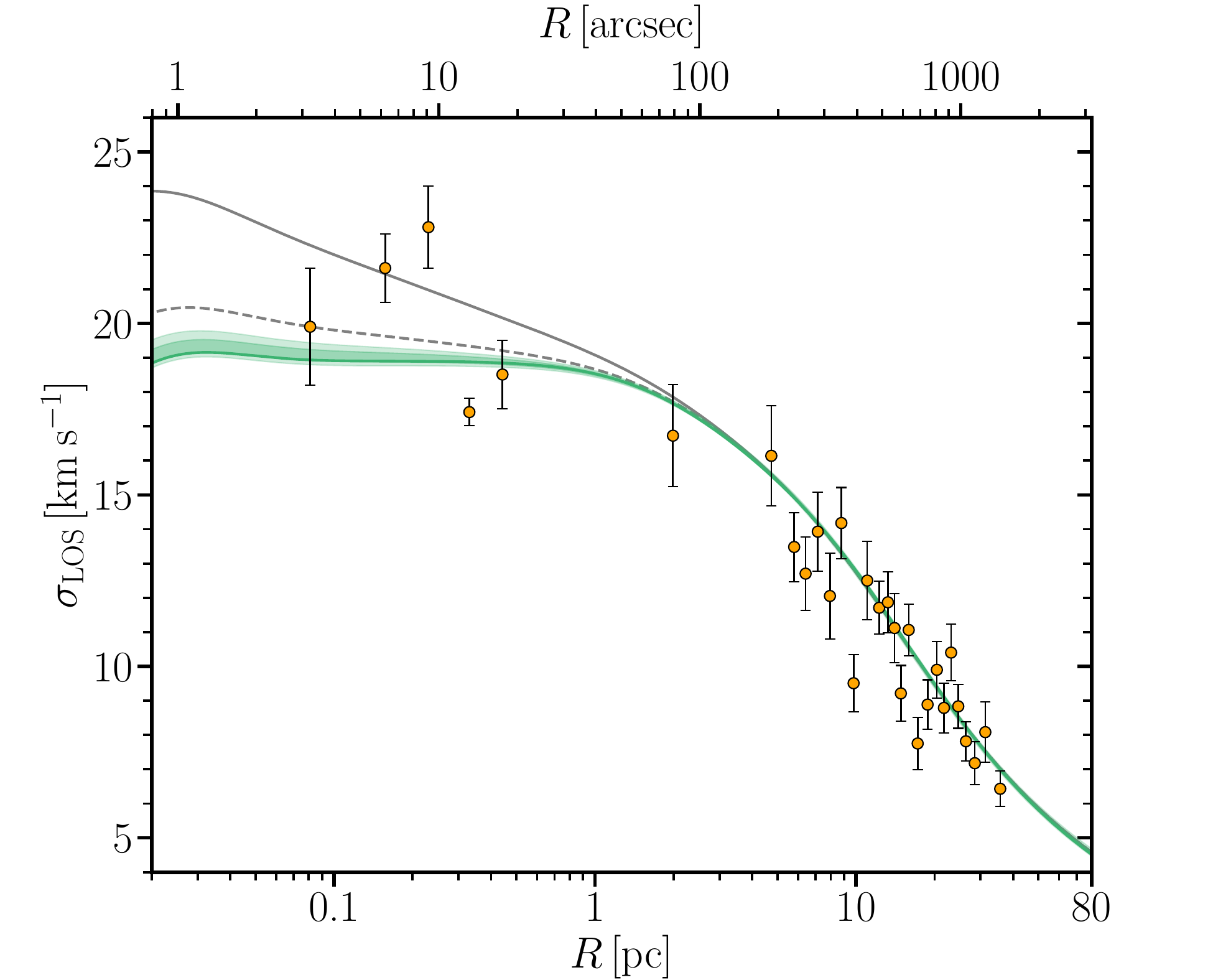}

\includegraphics[width=6.45cm]{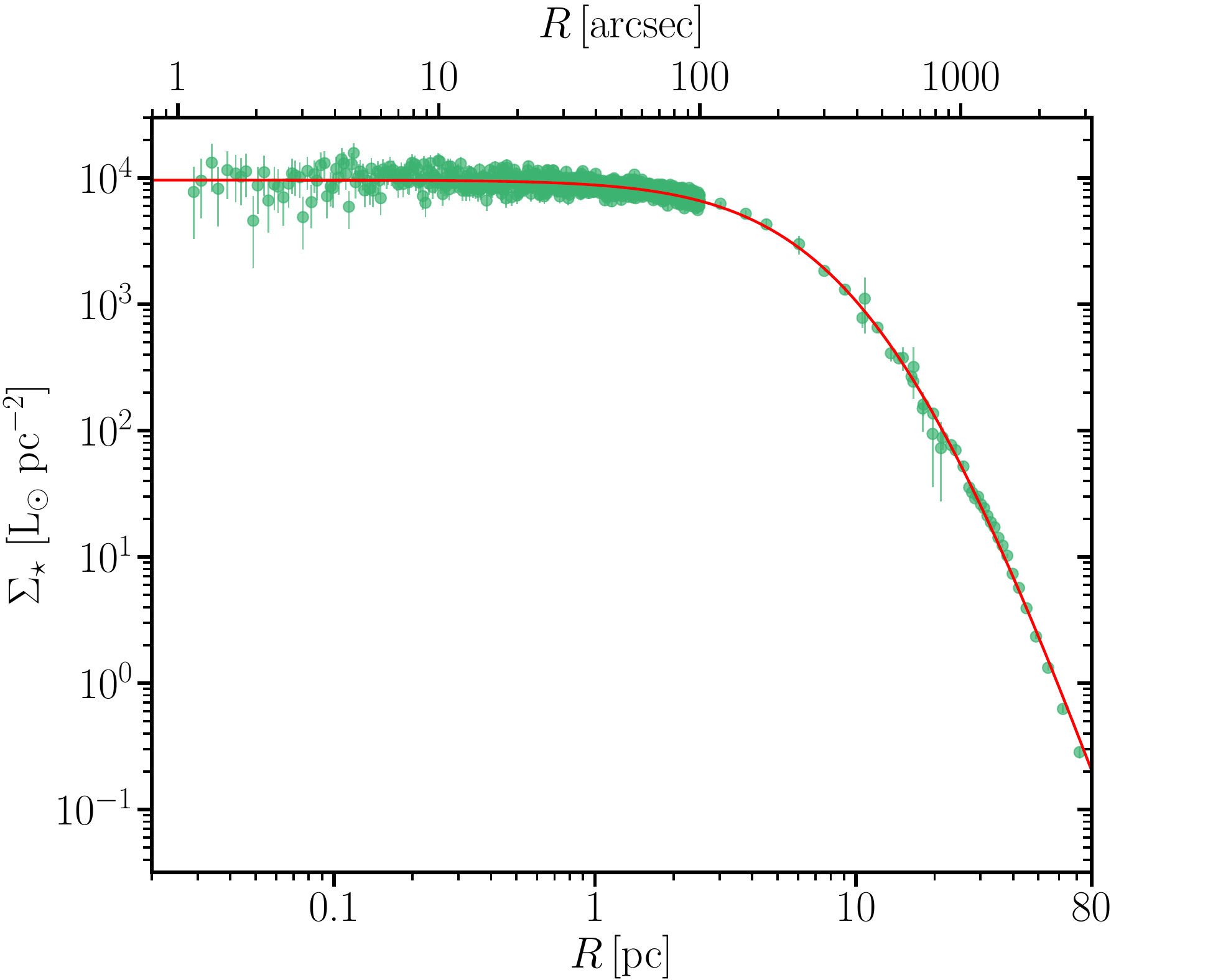}
\hspace{-0.7cm}
\includegraphics[width=6.45cm]{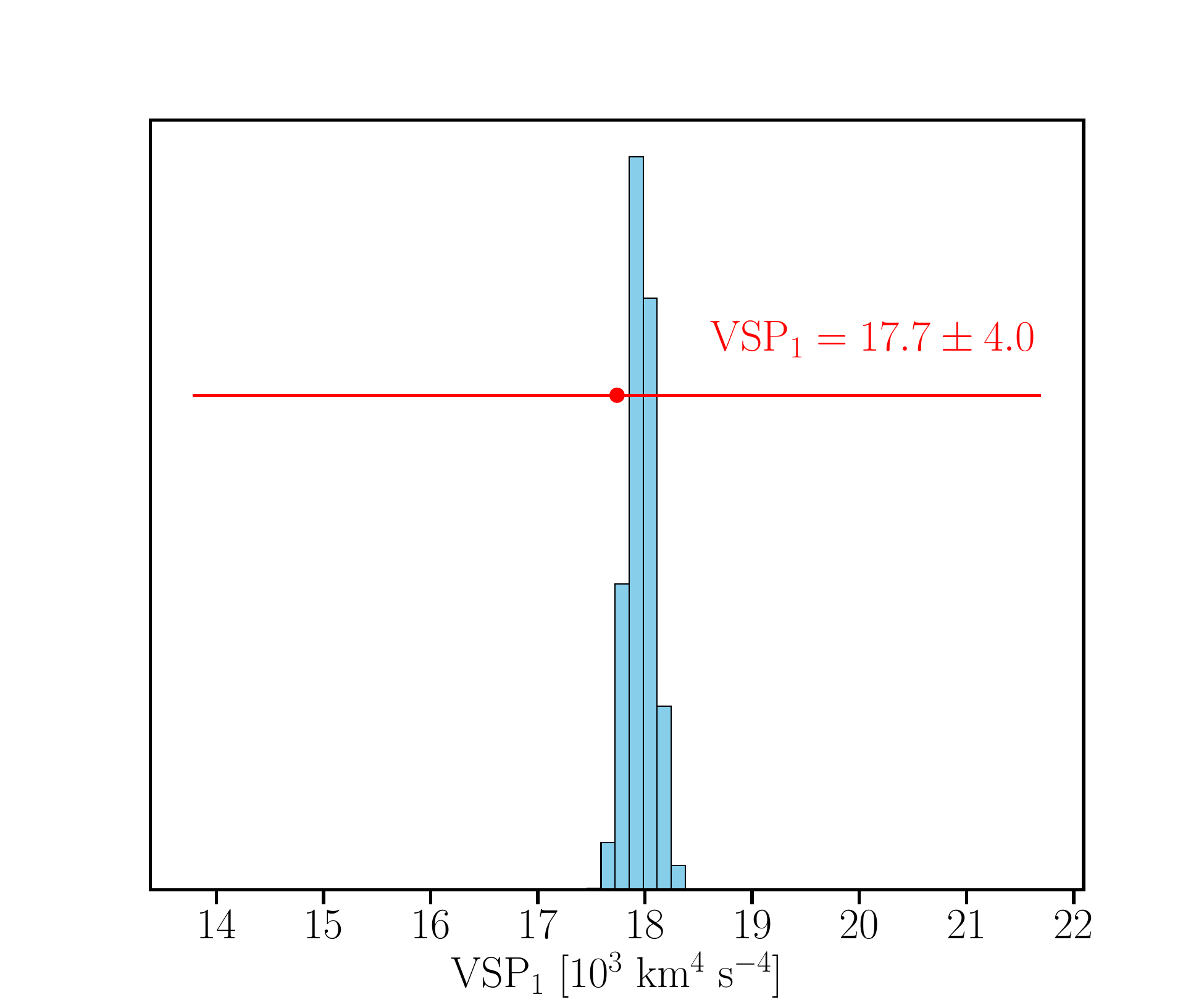}
\hspace{-0.7cm}
\includegraphics[width=6.45cm]{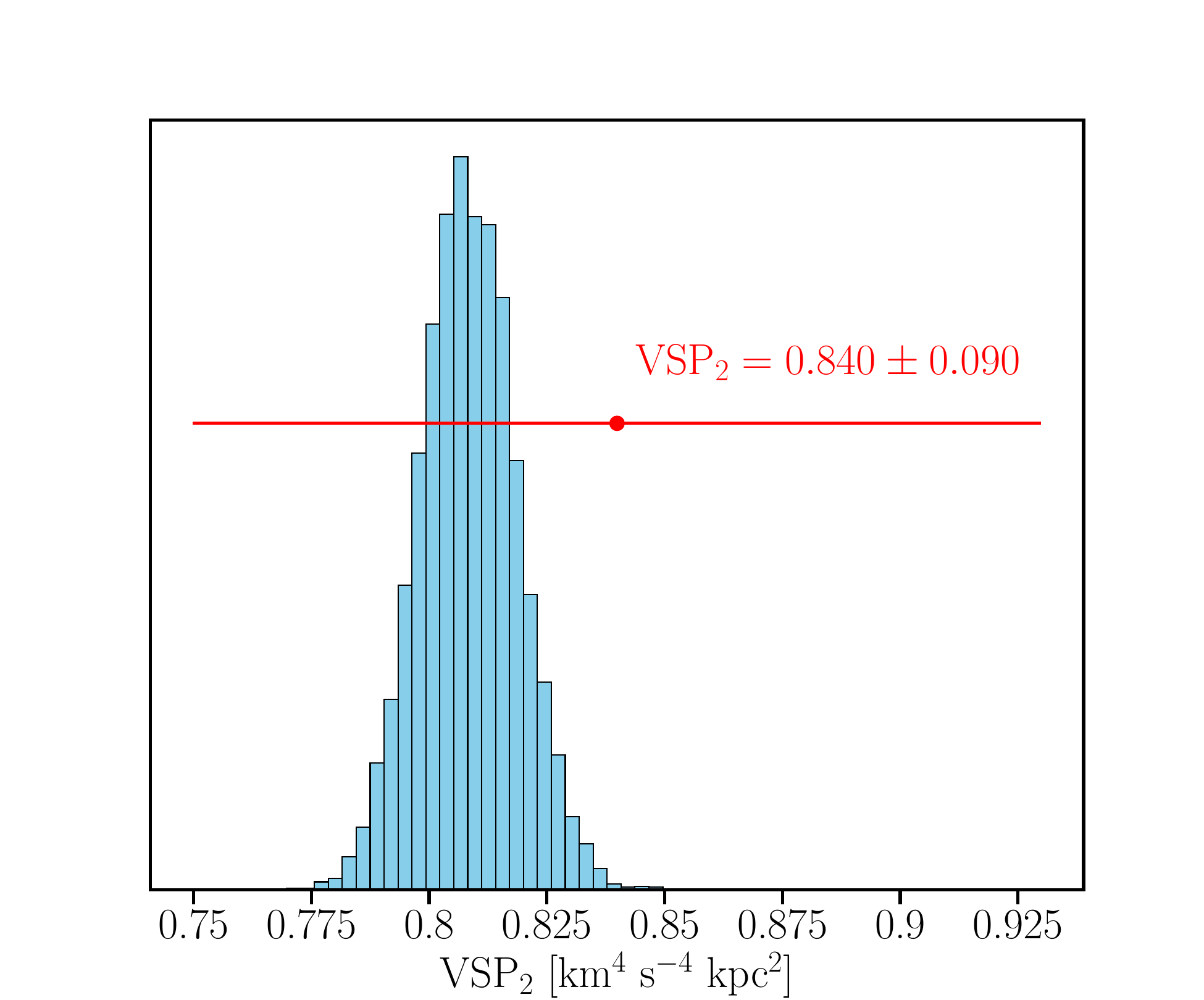}


 \caption{Fits to stellar kinematic observables and the photometric light profile. The profiles are a function of projected radius at the assumed distance of $5.2$~kpc. The green lines and bands correspond to the maximum posterior values, and the 68\% and 95\% CL centered at the median (respectively). For comparison, the gray (gray-dashed) line indicates the velocity dispersion obtained when a fraction of the dark component in the maximum posterior result is concentrated in the form of a $40,000$ ($10,000$) $\MS$ IMBH. Such IMBH models are strongly disfavored in our analysis. The stellar stellar kinematics datasets are described in  Sec.~\ref{sec:stardata}. All data, including the photometric profile, have been self-consistently binned with the center from \cite{anderson2010new}.   
 \emph{Upper left:}~Tangential proper motion velocity dispersion. \emph{Upper middle:} Radial proper motion velocity dispersion. 
\emph{Upper right:} LOS velocity dispersion.
\emph{Lower left:} Surface brightness profile used to determine the photometric component of the distribution. The red line shows the best-fit profile that was used throughout the analysis (see Sec.~\ref{sec:overview_method} for details). For presentation purposes, this profile has been normalized to match the central luminosity density of \cite{1995AJ....109..218T} after correcting for extinction using the same approach as in \cite{2017MNRAS.468.4429Z} with the reddening reported in \cite{1996AJ....112.1487H} (2010 edition) catalog.
\emph{Lower middle:} Posterior distribution for the virial shape parameter 1 as a result of the fits, with the red data point indicating the value with 1$\sigma$ errors computed by \texttt{binulator} (see Sec.~\ref{sec:stardata}).
\emph{Lower right:} The same as the lower left figure, but for the virial shape parameter 2.
 \label{fig:results}}
\end{figure*}

\begin{figure*}[h!]
    \centering
\includegraphics[width=7.5cm]{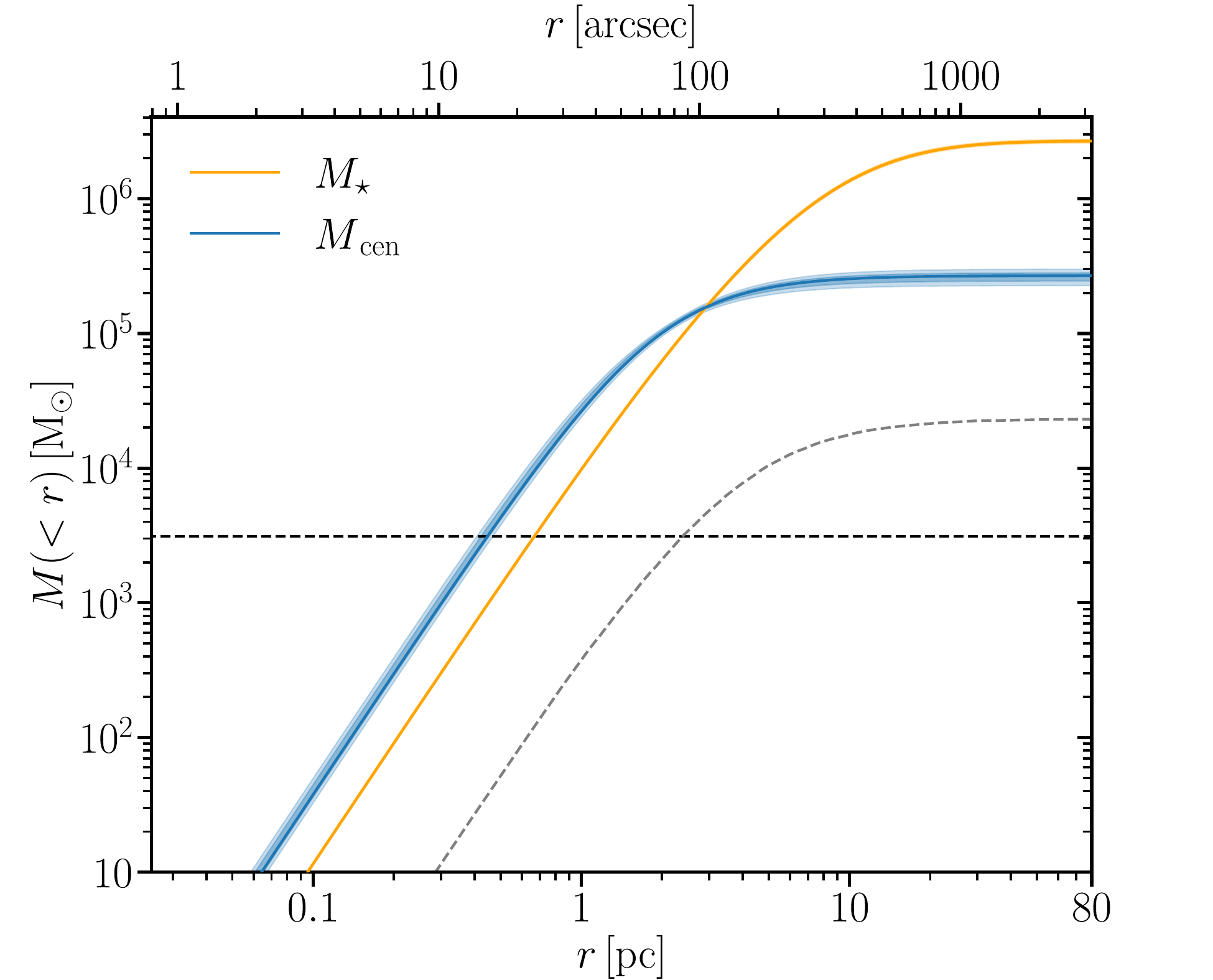}
\hspace{0.5cm}
\includegraphics[width=7.5cm]{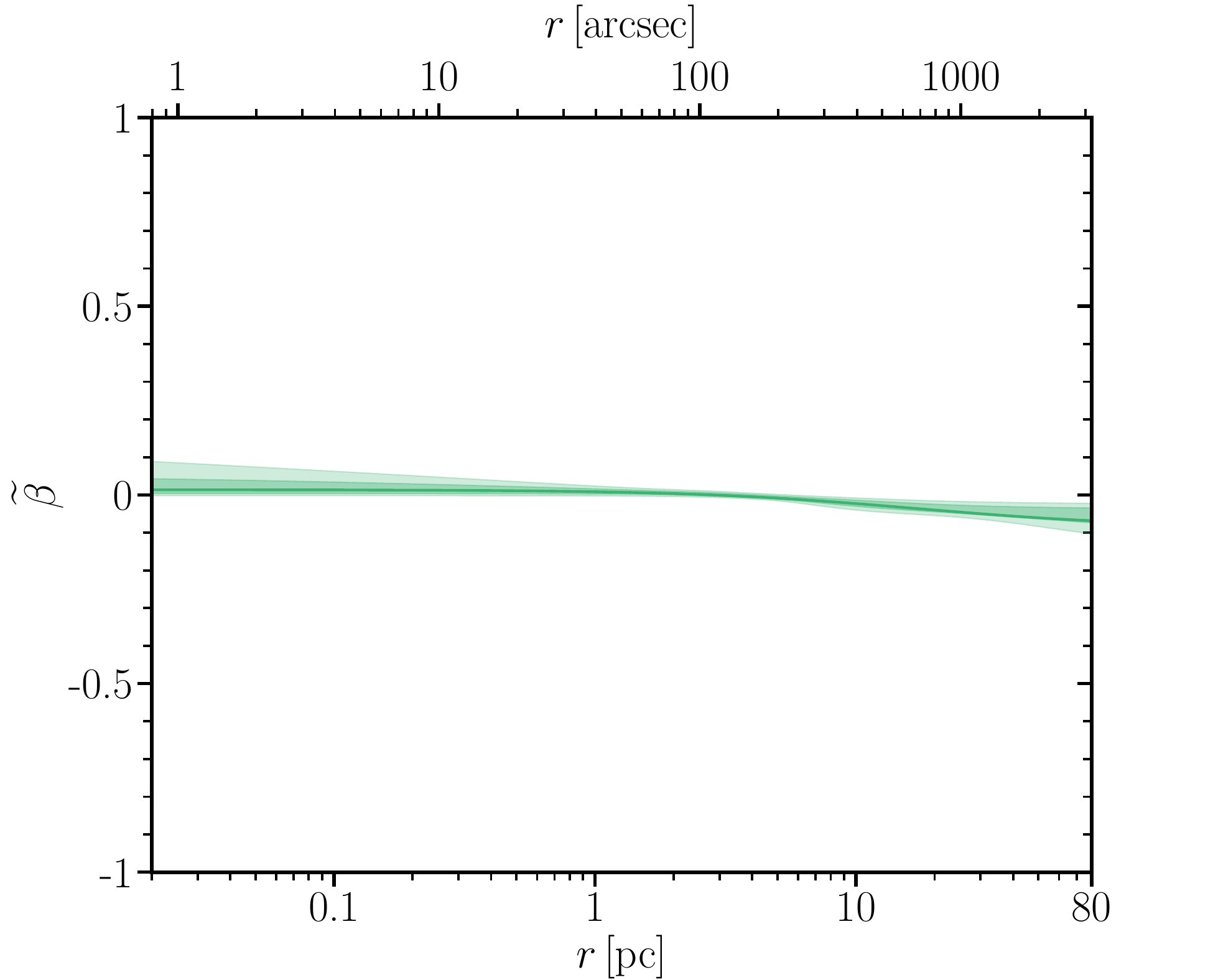}
\includegraphics[width=7.5cm]{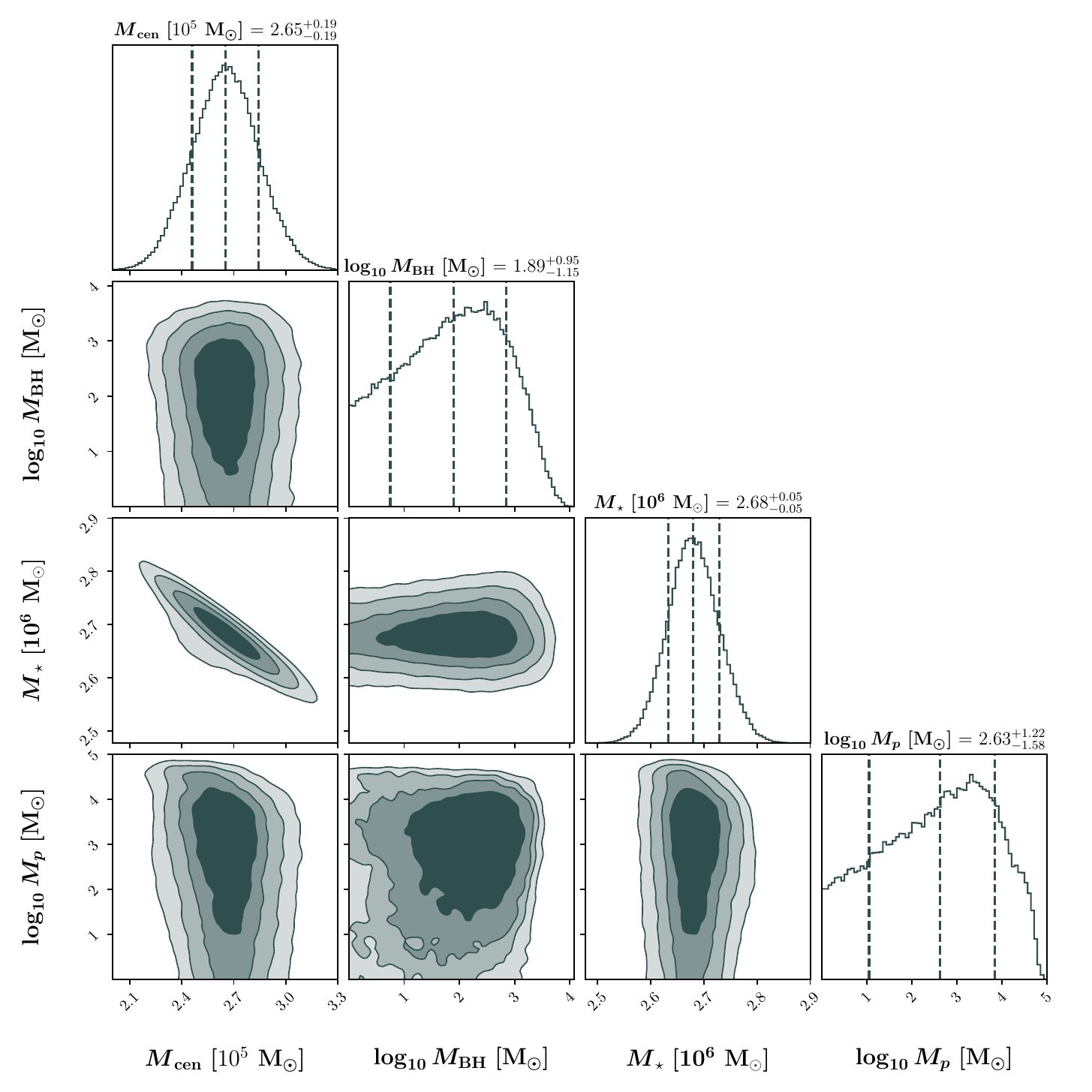} 
\hspace{0.5cm}
\includegraphics[width=7.5cm]{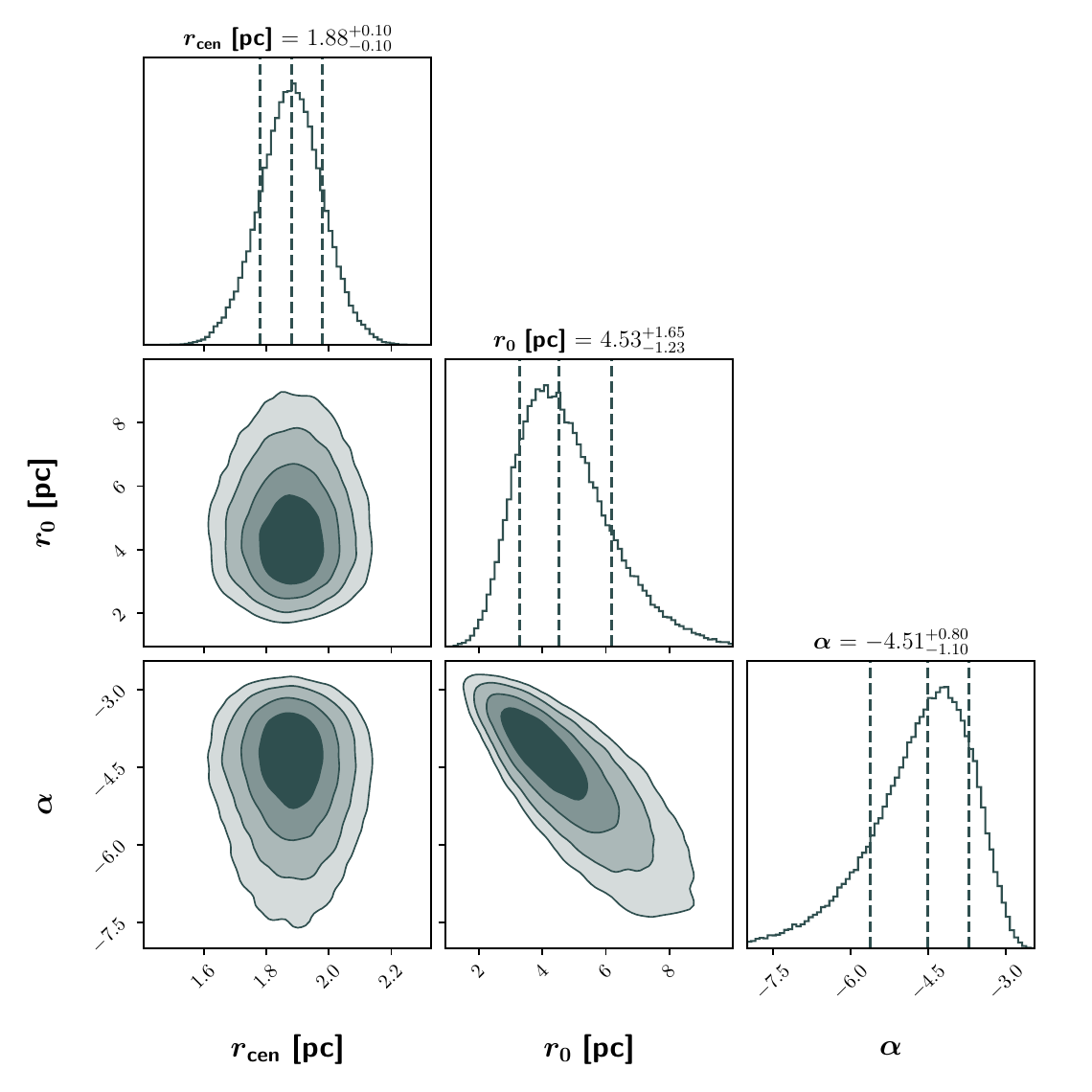}

 \caption{\emph{Upper left:} 3D enclosed mass profile including photometric, central remnants, black hole, and MSP mass components. The solid lines correspond to the maximum posterior with 68\% and 95\% CL regions. The black and gray dashed lines indicate the upper limit of the 95\% CL region for the black hole and MSP components, respectively. 
 \emph{Upper right:} Symmetrized anisotropy profile including the maximum posterior and 68\% and 95\% CL regions. The posterior distributions for the anisotropy parameters are presented in Appendix~\ref{app:postani}. The profile is close to isotropic, with a mild radial (tangential) anisotropy in the inner (outer) regions. This translates to the $1\sigma$ CL band being well within $|\widetilde{\beta}| < 0.1,$ or $|\beta| \lesssim 0.15$ for the vast majority of the range covered.
\emph{Lower left:} Posterior distributions for the masses of the various fitted components of our analysis, indicating the median and $68\%$ CL region at the top of each distribution.
\emph{Lower right:} Posterior distributions for the morphological parameters of the fitted mass profiles with respective median and  $68\%$ CL regions.  
 \label{fig:masani} 
 }
\end{figure*}

Fig.~\ref{fig:results} summarizes the fitting results obtained for the stellar kinematics and photometry observables used for the analysis.
For the LOS and PM velocity dispersions and the two virial shape parameters, we obtain a best-fit reduced chi-square value of $\chi_{\nu}^2 = 1.61$ for our 4-parameter anisotropy profile and a 3-parameter mass model (stellar mass and Plummer mass and scale radius). Note that we have excluded both the black hole and MSP profile components due to the fact they do not have relevant kinematic contributions in our fits and adding them does not materially improve them.\footnote{More precisely, the introduction of MSP and IMBH components leads to a higher $\chi_{\nu}^2$ value, due to the introduction of 4 additional parameters and a negligible effect on the total chi-square statistic, showing no statistically significant preference for these two components based on the additional degrees of freedom that they introduce.} This is consistent with our finding that the data strongly favor an extended central mass in the inner regions and a stellar/photometric component in the outer ones, while excluding at 3$\sigma$ CL an IMBH greater than $6 \times 10^3 \: \MS,$ or a kinematically relevant distribution of intermediate concentration traced by the MSPs. This bound on a putative IMBH mass is significantly more constraining than those reported in previous analyses. This result is illustrated in Fig.~\ref{fig:masani}, which shows the posterior distributions of the mass model parameters and the mass and anisotropy profiles. 

One reason for our bound being comparatively stringent is due to the extended central distribution that is favored, 
 limiting the kinematic contribution of additional central components. This result was found to be statistically significant, since the $\sim 10^4 \: \MS$ and $\sim 10^5 \: \MS$ total mass values required for kinematically signficant IMBH and MSP profile contributions (respectively) exceed the 3$\sigma$ limits of our derived posterior distributions, and are thus self-consistently excluded by the fit. We also found that, when fitting only the MSP component to model extended remnants, the MSP distribution was forced to emulate the concentration of our central mass  favored by the kinematics, but failed to reproduce the more extended distribution of the MSPs, which is implicitly accounted for in the position likelihood components. This can be seen when comparing the cumulative distribution functions of the various components, which are examined in Sec.~\ref{sec:mspan} (Fig.~\ref{fig:acproj}). 
This result indicates the need for a more concentrated mass component that is distinct from one traced by the pulsars. 


Our results show a clear preference for a two-component model with an extended central mass of $\sim 2 - 3 \times 10^5 \: \MS$ and scale radius between $1.5$ and $2.2$ pc at the 3$\sigma$ level. This is favored over an IMBH, leading to a 3$\sigma$ upper bound of $6 \times 10^3 \: \MS$ 
and placing a coexistence region for a putative IMBH of at most a few thousand solar masses. Our analysis also places a $3\sigma$ upper bound for the MSP distribution of $6\times 10^4\: \MS,$ strongly limiting the kinematic contribution of the more extended MSP component.\footnote{This result can be more dependent on the modeling of the other extended components in the distribution, especially if these exhibit degeneracies with each other. However, this is not the case in our analysis, as a kinematically relevant contribution is disfavored for this component. Further, we found that the precise values the $3 \sigma$ bounds in this study can show  moderate stochastic variability when comparing multiple independent runs of the fits, while this has only a moderate effect (agreeing at the specified number of significant figures in most cases), we chose the cited values conservatively so that they exclude \emph{at least} at the $3\sigma$ level, across the multiple (more than 10) runs considered.} 

This is not present in previous analyses by construction, since these fit a single component of either a point mass or an extended distribution accounting for remnants, but not both simultaneously.\footnote{Note also that while a single-component Plummer sphere or a similar extended mass model can in principle accommodate for a point mass in the limit of vanishing scale radius, it suffers the limitation of being a mutually exclusive model in that it does not explore the coexistence between a point mass \emph{and} an extended one.} While previous studies have indicated the kinematic degeneracy between extended central distributions and an IMBH, in our analysis we are able to consider them simultaneously in a self-consistent fit. 

One way the stellar kinematics favor this result is by having velocity dispersion profiles that are relatively flat in the central regions, with a value of $\lesssim 20$\,km/s within the inner $\sim 0.1$ pc,  while still being sufficiently elevated within a few pc from the center to favor a significant extended central mass component of $\sim 2-3 \times 10^5 \: \MS$ that is needed in addition to the photometric one. Large ($\gtrsim 10^4 \: \MS$) IMBH models predict distinct cusps in velocity dispersion profiles (cf. Figure 5 of \cite{2019MNRAS.482.4713Z}), while more extended distributions do not. The fact that no such cusp is observed in our data, which has sufficient resolution to probe this central region, disfavors the presence of such component, and is instead consistent with the flatter profile predicted by the extended central mass component.

The similarity between the three velocity dispersion components explored is a manifestation of the approximate isotropy of the distribution, leading to a relatively well-constrained anisotropy profile. This can be observed clearly in Fig.~\ref{fig:centers}, where the three velocity dispersion components are plotted in the same figure, and is also apparent from the (symmetrized) anisotropy profile in Fig.~\ref{fig:masani}.


The dynamical mass of the photometric component yields $M_{\star} =  2.68 \pm 0.05 \times 10^6 \: \MS, $ which for a total luminosity of $\sim 1.25 \times 10^6 \: L_{\odot}$ would imply a mass-to-light ratio that is comparable, but arguably higher than those typical of stellar populations, showing consistency with previous studies, with some suggesting $\OC$ to have high remnant fractions, for instance, in the form of white dwarfs \citep{2024MNRAS.529..331D}. For the photometric profile in Fig.~\ref{fig:results} (middle right), we found this to be very well constrained by the data in morphology via the surface brightness data, yielding a very good fit with $\chi_{\nu}^2 = 1.13$ for the 4-parameter model used. This, in addition to previous runs we performed where we fitted it with the kinematic data, justifies the simpler approach to fix the profile to its best-fit values.

\begin{figure}[t]
  \centering
 \includegraphics[width=\linewidth, keepaspectratio]{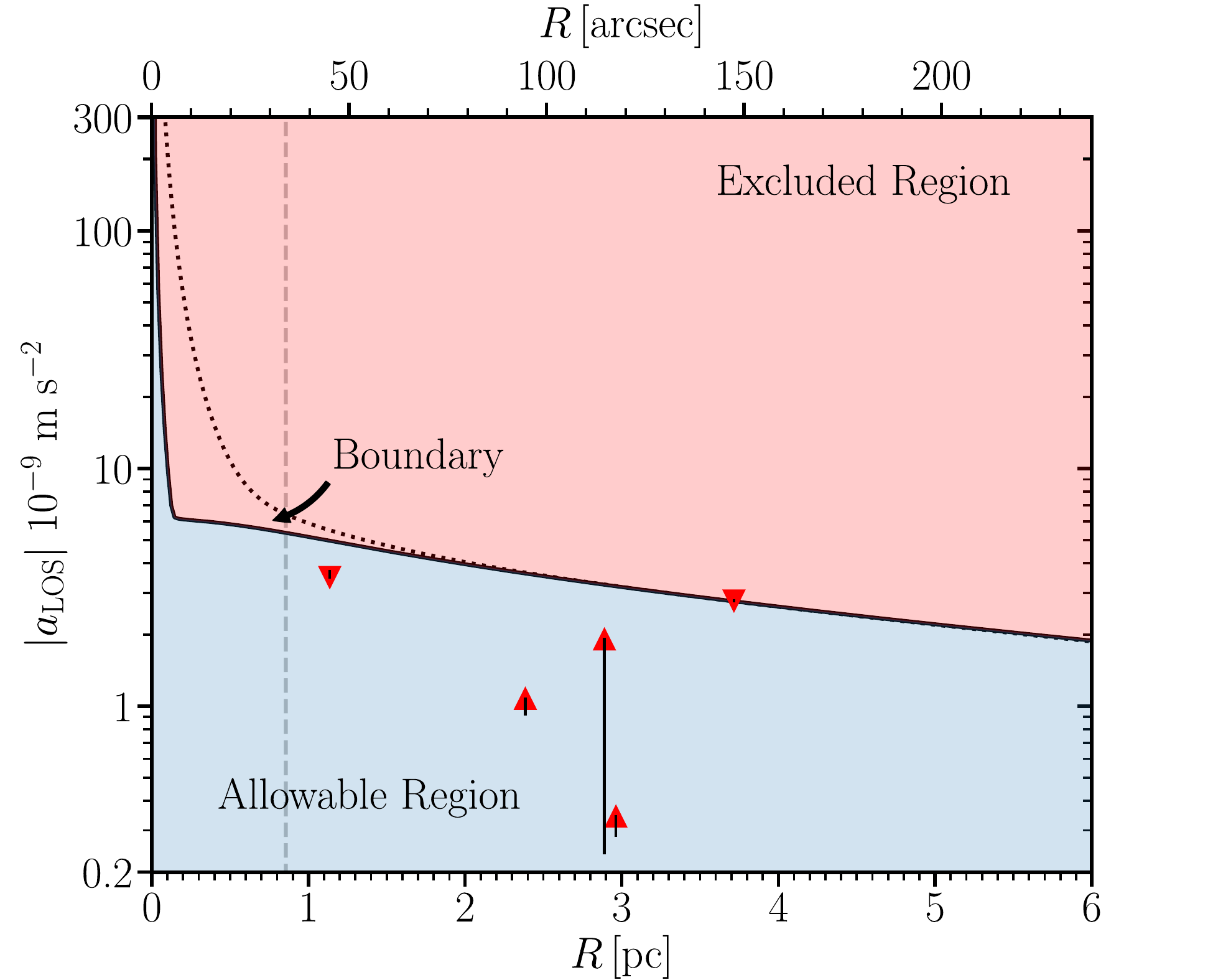}
 \caption{Absolute values of MSP LOS accelerations. The black, solid line denotes the 95\%CL upper bounds as inferred from the mass profiles of the posterior distribution. This region marks the boundary between the excluded region (red) and the allowable one (blue), where it is possible to find a value of $l$ (i.e. a position within the cluster) compatible with the LOS acceleration at a given projected radius. The upwards (downwards) pointing triangles denote the observed MSP accelerations with positive (negative) values, as inferred from the MSP timing data. Errors are not included as they fall below the size of the data points. The vertical black lines denote the intrinsic spin-down components leading to the intrinsic LOS accelerations that trace the GC potential, based on the 68\% CL regions in the magnetic field strength posteriors. This interval accounts for 84\% of the posterior distribution as it includes the lower tail below the 16\% percentile value in addition to the 68\% CL interval contributions.
 The black, dotted line shows the corresponding bound for an illustrative model where $\sim 15$~\% of the central mass distribution is concentrated in the form of a $4\times10^4 \: \MS$ IMBH.
 As can be seen, this model is still consistent with the pulsar accelerations (triangles), but is in tension with the proper motion and line of sight velocity data. A lower mass IMBH is still allowed (see the central allowable region that reaches to higher accelerations). The gray, dashed line indicates the projected position of the innermost detected MSP in $\OC$ at $\sim 0.86$~pc from its center. 
 \label{fig:accpul}}
\end{figure}


Fig.~\ref{fig:accpul} illustrates the fits to the MSP kinematics corresponding to the LOS acceleration profiles for pulsars for the 5 MSPs with suitable timing solutions, noting that for the 13 remaining ones only projected positions were used as a constraint on the MSP distribution profile. We may now ask, more generally, the degree to which the MSP data are constraining of the overall mass distribution in this analysis. An important result in this regard is that our self-consistent fit with the MSP data yields a central mass component that is systematically more extended and massive when MSP accelerations are included, with in increase of $\sim 20 \: \%$ in both the mass and scale radius of the distribution, with the black-hole and photometric components being comparatively unaffected (cf. Appendix~\ref{app:noaf}, where a fit is performed without including MSP accelerations). 

The dominant contribution to this effect appears to be attributable to the most distant MSP acceleration data point, which effectively sets a lower limit on the enclosed mass at that point. The other data points were found to be independently consistent with the remaining pulsars, including the innermost one, whose relatively high acceleration was found to be potentially conflicting by \cite{Dai:2023pzr} with the models they explored, as these do not account for the presence of an inner dark mass such as the one favored in our analysis. \cite{Dai:2023pzr} also indicated a potential tension with this outermost point during their analysis. However, the degree to which such a tension is present in our analysis is necessarily dependent on the assumptions made on the intrinsic spin-down component, due to the fact that the lower bound implied by the observed apparent acceleration is within the allowable region of our analysis. While this data point favors a low intrinsic spin-down component, we found it to agree at the 2$\sigma$ level with the observationally derived prior on the magnetic field strength parameter, with the other MSPs agreeing at the 1$\sigma$ level.  Considering also the significant observational and modeling uncertainties inherent to intrinsic spin-down determinations, we do not consider this to imply a significant tension in our analysis. It is interesting to note, however, that the constraining power effected by the outermost data point is robust with respect to intrinsic spin-down effects, as higher negative contributions to the acceleration would only increase the degree to which a more massive and extended distribution is favored. 

In summary, we find the tension examined by \cite{Dai:2023pzr} to be alleviated by this analysis as a result of including a central dark mass component in our model which emulates an extended cluster of stellar remnants (see discussion in Sec.~\ref{sec:rem}), and by our statistical treatment of intrinsic spin-down modeling which allows quantifying the significance of putative tensions. 

For illustrative purposes, we also show in Fig.~\ref{fig:accpul} the effect of concentrating $\sim 15 \%$ of the central mass component in the form a $4\times10^4\: \MS$ IMBH, which is representative of some of the values in the literature \citep{Noyola:2008kt, 2017MNRAS.464.2174B}. 
We observe that, while in both models extremal LOS accelerations are allowable, the larger IMBH model would produce significant extremal accelerations over a larger region within the inner $\sim 0.5$ pc, comparable to the radius of the circle of influence $r_{\rm inf} \sim $ of the putative IMBH. A detection of such an extremal acceleration within this region would constitute a `smoking-gun' signature of such an IMBH. Such scenario is disfavored by stellar kinematics in our analysis, except for the very inner region within $\lesssim 0.15$ pc, where a smaller IMBH of at most a few thousand solar masses is still allowable.\footnote{We determine this region as that for which our analysis allows a measurably large extremal acceleration that could no be accounted by the extended components alone. This corresponds to the blue, thin inner cusp seen in Fig.~\ref{fig:accpul}.} This  limited presence of extremal accelerations remains a falsifiable prediction in the context of our analysis, should new MSPs be detected in the innermost regions of $\OC$ (cf. gray, dashed line of Fig.~\ref{fig:accpul}). 




\section{Analysis of the pulsar distribution}
\label{sec:mspan}


\begin{figure}[t]
    \centering
\includegraphics[width = \linewidth, keepaspectratio]{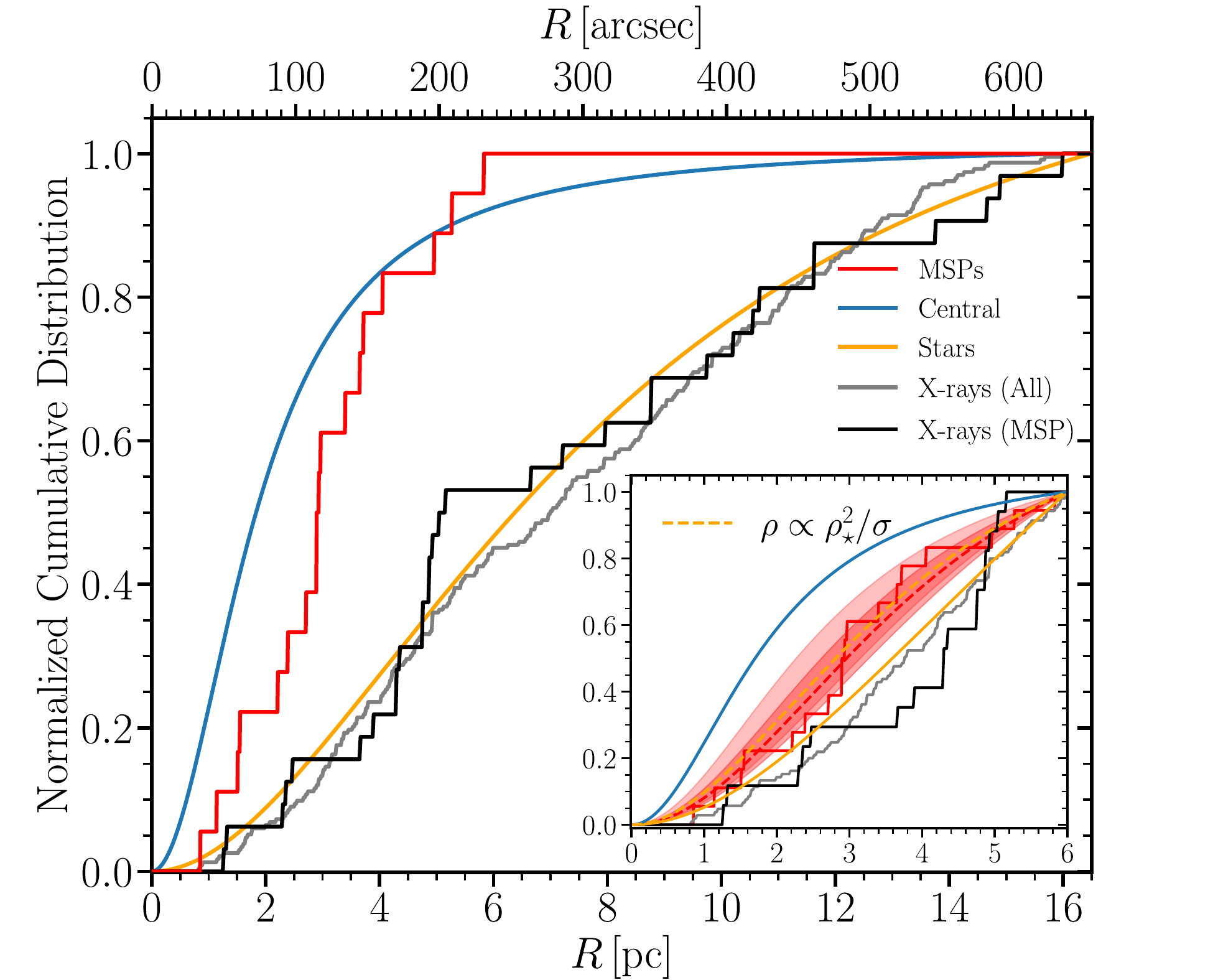}
\caption{Normalized cumulative distributions of MSPs (red), the more concentrated central mass emulating heavier stellar remnants (blue), stars (yellow) and X-ray emitting sources (black and gray) in $\OC$ as a function of projected radius. The inset shows a close-up view of the distribution normalized at a smaller radius, where all of the 18 MSPs are located. The inset includes the 68\% and 95\% CL regions of the MSP profile fit (shaded bands). The red dashed line denotes the median of the inferred cumulative distribution from the MSP 3D density from Eq.~\eqref{eq:pul_3d}. The dashed-orange line indicates the predicted distribution derived from the stellar encounter rate that provides a remarkable match to the MSP distribution. We also show counts of X-ray sources observed in $\OC$ studied by \cite{2018MNRAS.479.2834H},
 showing the total count of 233 objects (black) and a subset of $\sim 32$ of the objects that share luminosities and X-ray colors compatible with known MSPs from other GCs (gray), as presented in Figure 10 of \cite{2005ApJ...625..796H} (see also the discussion in \cite{2018MNRAS.479.2834H}).
Over the radial range of the inset figure, this count is reduced to 105 and $\sim 17$ sources, respectively.
\label{fig:acproj}}
\end{figure}


\begin{figure}[t]
    \centering
\includegraphics[width=\linewidth, keepaspectratio]{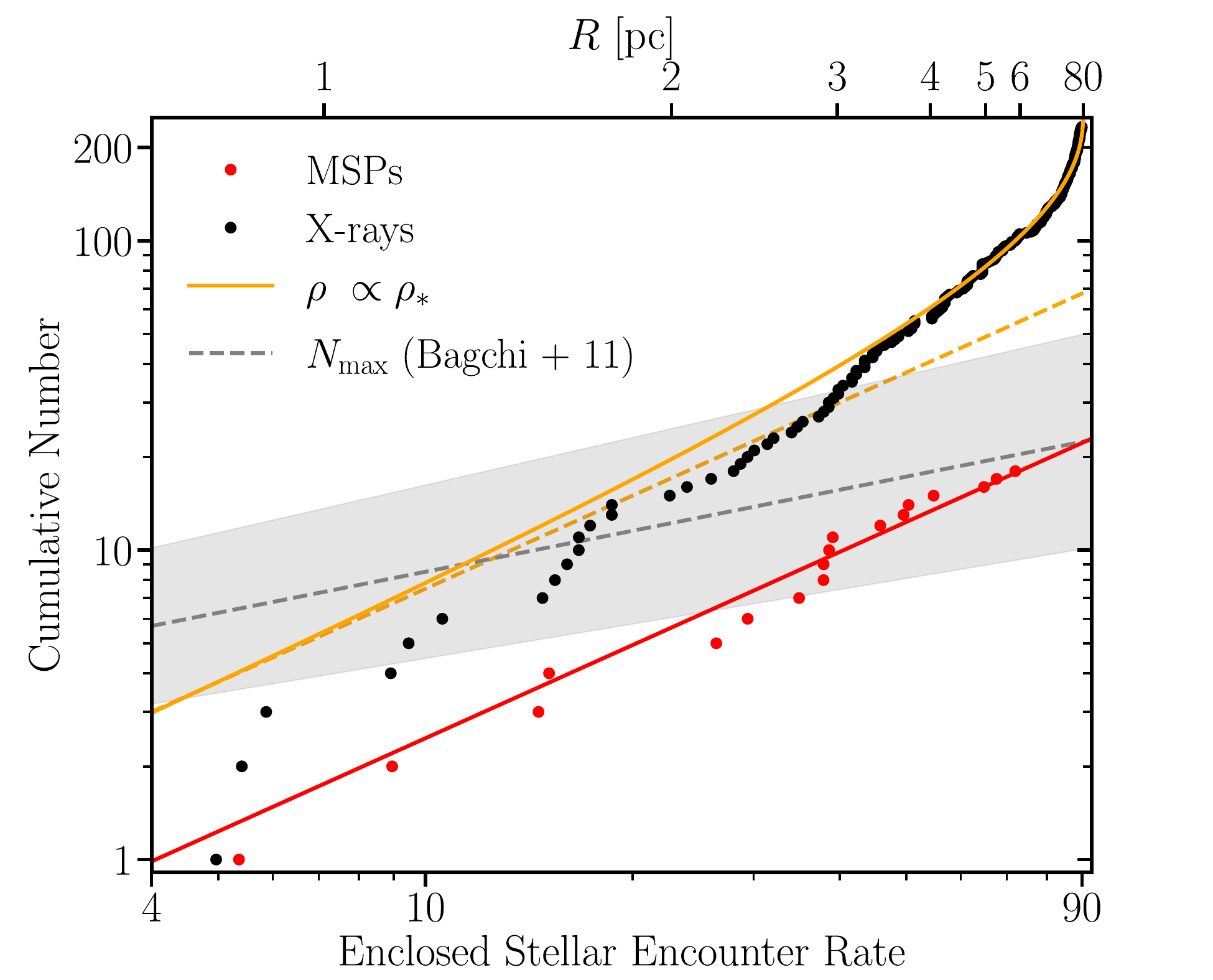}
\caption{MSP abundances as a function of stellar encounter rates. The red points are the observed cumulative number of MSPs as a function of the enclosed encounter rate at a projected radius $R$ (upper x-axis). The red line corresponds to a least-squares linear fit, showing a clear linear dependence. The black points denote the same quantities for the full list of 233 X-ray sources from \cite{2018MNRAS.479.2834H}, while the orange line corresponds to the stellar distribution, normalized to match these sources. These show a distinct, non-linear, dependence, with encounter rates that are not followed by the MSPs. The orange, dashed line is a linear extrapolation shown for comparison.
Lastly, the gray dashed line indicates an upper estimate on the total number of MSPs as a function of total encounter rates of GCs, with $1\sigma$ error bands. This is based on the parametric fit performed by \cite{2022MNRAS.511.5964Z} using the phenomenological estimates of \cite{2011MNRAS.418..477B}, based on luminosity functions, and assuming the updated total stellar encounter rates of \cite{2013ApJ...766..136B}. Units for $\OC$'s total encounter have been normalized to match the 90.4 central value of \cite{2013ApJ...766..136B}.
\label{fig:ser}} 
\end{figure}

Fig.~\ref{fig:acproj} shows the (normalized) cumulative distributions of the MSP, photometric, and extended central mass components derived from our model fits. We also plot an independent dataset of X-ray sources shown for comparison purposes. The morphology of the MSP distribution is well constrained by the MSP data. It is clear from the figure that the MSP distribution is significantly more concentrated than the photometric component. The X-ray dataset was introduced to compare whether it may be independently traced by the components explored in this study. This is of particular interest given that 11 of the 18 known MSPs in $\OC$ possess X-ray counterparts \citep{2023MNRAS.526.2736Z, Dai:2023pzr}. While it seems highly plausible that some of the MSPs are in fact found among these X-ray sources, interestingly, we find that the aggregate distribution of X-ray sources appears to be traced by the (more extended) stellar density profile, which, as implied by our previous discussion, was determined independently from stellar photometric data. This result appears to be essentially independent of which of the two X-ray sources is included, with the one sharing MSP-like properties (black) matching the total distribution (gray). 


We now consider what the more concentrated MSP profile that we find here can tell us about MSP formation. MSPs have long been thought to descend from low mass X-ray binaries (e.g. see \cite{2013ApJ...766..136B, 2014A&A...561A..11V}). These form dynamically in GCs at a rate set, to leading order, by the rate of stellar encounters.\footnote{This may also apply more generally to comparable models involving formation via two-body encounters that are not limited to low mass X-ray binaries.} 
Such models predict that the abundance of MSPs in GCs should scale with the (total) stellar encounter rate for the cluster $\Gamma \propto \int dr \: r^2 \rho_{\star}(r)^2 / \sigma(r),$ where $\rho_{\star}(r)$ is the 3D density profile of stars, $\sigma(r)$ is the 3D velocity dispersion profile of the stars, with $\rho_{\star}(r)^2/\sigma(r)$ being the encounter rate density, which encapsulates the rate of two-body stellar interactions (hence the $\rho_\star^2$) and is enhanced by gravitational focusing (hence the inverse relation to $\sigma$). In practice, however, we may expect some deviation from this relation as the formation of MSPs depends on the binary stellar encounter rate that can differ from the single-star one \citep[e.g.][]{2014A&A...561A..11V}, binary population statistics, loss of stars from the cluster (impacted by the cluster escape velocity; \citealt{2024arXiv240518228Y}), neutron star retention fractions (e.g. \cite{2014A&A...561A..11V}), and the effects of mass segregation (e.g. see \cite{2013ApJ...766..136B} and references therein). Despite these caveats, numerous studies have reported a correlation between the MSP population abundances and the (total) stellar encounter rate in GCs \citep{2013ApJ...766..136B,2022MNRAS.511.5964Z}.

Previous studies have focused on establishing correlations between total encounter rates and abundances of MSPs in GCs. 
Following the above discussion, we decided to extend these analyses by assessing whether a similar relation exists at the intra-cluster level. In particular, we explore whether the MSP population density scales with the stellar encounter rate density so that $\rho_p (r) \propto \rho_{\star}(r)^2 / \sigma(r),$ which would naturally predict a linear correlation with the total encounter rate when both of these quantities are integrated over the volume of the cluster.

In the inset panel of Fig.~\ref{fig:acproj}, we compare our derived radial density profile of MSPs (shaded bands) with a model in which the MSP distribution depends solely on the stellar encounter rate (yellow dashed line). This gives an excellent match. 
Indeed, if we perform an indicative fit of the form $\rho_{p}(r) \propto \rho_{\star}(r)^\gamma /\sigma(r),$  we find that $\gamma = 1.9 \pm 0.3,$  showing excellent agreement for the central value of the inferred MSP distribution and well within its 68\% CL bounds. Since we found that $\sigma(r) \propto \rho_{\star}(r)^{0.2}$ to a good approximation for the region covered by the inset, this relation can also be expressed directly in terms of the stellar distribution as $ \rho_{\star}(r)^{\gamma} / \sigma(r)  \propto \rho_{\star}(r)^{\gamma - 0.2},$  so that $ \rho_p(r) \propto \rho_\star(r)^{1.7}$ for the central values.

Fig.~\ref{fig:ser} shows explicitly the dependence of the distributions considered as a function of the encounter rates, which for the case of $\OC$ can be mapped to the profile via the enclosed encounter rate over a cylindrical volume with projected radius $R$, which we define as:
\begin{equation}
\label{eq:encenc}
\Gamma (<R)  \propto \int_{-\infty}^{\infty} dl \int_{0}^{R} dR' R' \rho_{\star}(r)^2 / \sigma(r),
\end{equation}
where $r = \sqrt{R'^2 + l^2}$.

Encounter rates are usually expressed in arbitrary units, which in our case have adapted to be consistent with \cite{2013ApJ...766..136B} for the total value (i.e. $\Gamma(<\infty) \sim 90.4$). For obtaining the constant of proportionality $K$ mapping the encounter rate  to the number of MSPs, we perform a least-squares fit (Fig.~\ref{fig:ser}: red line) where we model the cumulative number of MSPs as: 
\begin{equation}
\label{eq:fitser}
\text{Number  of MSPs} \: (<R) = K \: \Gamma (< R), 
\end{equation}
yielding $K \sim 0.25.$ This closely matches the central value of the total number of MSPs ($\sim 23$) independently predicted by the phenomenological relation from \cite{2022MNRAS.511.5964Z} (Fig.~\ref{fig:ser}: gray, dashed line). However, the $1\sigma$ upper bound of this relation implies more than double the abundance, suggesting that $\OC$ could host a significant number of yet-undiscovered MSPs. One can also use the kinematic component of the MSP distribution to derive a strict upper bound on the MSP abundance assuming the totality of the dynamical mass being due to $\sim 1.6 \; \MS$ MSP binary systems, obtaining $\sim 4 \times 10^4 $ at $3 \sigma$ CL (c.f. Fig.~\ref{fig:masani}).

The most significant result from Fig.~\ref{fig:ser} is the clear linear scaling observed for the MSP distribution, yielding a relation analogous to those from previous works for total encounter rates and abundances of GCs, but for the enclosed intra-cluster populations instead. This validates the model from Eq.~\eqref{eq:fitser}, showing excellent consistency with the expectation from stellar encounter models  of MSP formation discussed earlier.

On the other hand, for the case of the stellar density profile and the X-ray sources independently traced by it, we observe markedly different behavior, with a distinct non-linear dependence followed in both cases. This further illustrates the differences between these distributions, with the linear scaling behavior being a unique characteristic of the MSP population.

To the knowledge of the authors, while such scaling relations have been explored at the level of total encounter rates and MSP counts for GC populations, this constitutes the first analysis where a similar relationship is validated as a function of radius within a single GC, thus motivating a functional form for the profile of the pulsar distribution within the cluster. 

It is also interesting to note that, following the discussion of Sec.~\ref{sec:rem}, since significant mass segregation for lighter stellar remnants (which MSPs are part of) is not expected for $\OC,$ a different mechanism would be needed to explain their more concentrated distribution. This further validates the stellar encounter model, which naturally predicts the rate of encounters to be proportional to the square of the stellar density, and thus leads to a more concentrated distribution. This consideration, in conjunction with its abundance of recently discovered pulsars, makes $\OC$ an ideal candidate to study such models of MSP formation.\footnote{Cf. \cite{Chen:2023lzp}, where the idealness of $\OC$ for probing the MSP distribution based on its lack of mass segregation was also independently argued as a promising direction for future studies.}

\section{Discussion}
\label{sec:discussion}
\subsection{Method}
\label{sec:meth}

The conflicting results in the literature regarding the nature of a central dark mass component in $\OC$ are likely attributable to systematic differences in the data considered and the mass-anisotropy modeling adopted. For instance, \cite{Noyola:2008kt} only considered a limited dataset of radial velocities for the stellar kinematics, while also assuming isotropy, which can lead to degenerate effects with anisotropic mass components (e.g. \cite{2017MNRAS.468.4429Z, 2017MNRAS.471.4541R}). As noted in Sec.~\ref{sec:rem}, degenerate effects with central remnants should also be accounted when considering the potential presence of IMBHs. These aspects have been extensively addressed in our analysis, with the inclusion of flexible mass-anisotropy modeling and a full self-consistent consideration multiple stellar kinematic data, including also MSP accelerations as an additional constraint.

\cite{2010ApJ...719L..60N} found a dependence of the inferred central mass with the assumed location of the kinematic center in their analysis, with the center used in \cite{anderson2010new} leading to a lower estimate of the inferred central mass. In a recent study from \cite{2024MNRAS.528.4941P}, employing 3 independent center determinations, it was concluded that the favored center led to inferred stellar kinematics determinations that were consistent with the results from different datasets across multiple studies, including those of \cite{anderson2010new} using HST data. However, a significant tension was found with the originally assumed center in \cite{Noyola:2008kt}. This is also consistent with the findings of \cite{anderson2010new}, where the \cite{Noyola:2008kt} center was ruled out at high confidence and determined to be susceptible to systematic errors not originally considered in the study of \cite{Noyola:2008kt}. 

We have addressed this issue using data with self-consistent centers (including our derived binning) for the kinematics  and photometry, while also considering different centers such as the ones from \cite{2010ApJ...719L..60N} and \cite{Noyola:2008kt}. These results are shown in Appendix~\ref{app:centers} (Figs.~\ref{fig:post_centers} and \ref{fig:centers}). While the inferred velocity dispersion profiles appear broadly consistent with those derived with the \cite{anderson2010new} center, we found that the central mass component derived with this center has a median mass (scale radius) that is $\sim 15 \%$ ($\sim 11 \%$) lower using the \cite{2010ApJ...719L..60N} center and  $\sim 14 \%$ ($\sim 10 \%$) higher using the  \cite{Noyola:2008kt} center, with less significant effects on the remainder of the mass components in both cases. This indicates a somewhat more extended inferred central mass distribution when using the \cite{2010ApJ...719L..60N} center and a more concentrated one for the \cite{Noyola:2008kt} center. This shows that the essential conclusions of our analysis are robust with respect to the choice of centers considered, with the main effect being a moderate variability in the concentration of the central mass component.

Another aspect in the literature concerns the various cluster distances adopted. We note that the close-to-isotropic distribution which is apparent in the data (cf. Figs.~\ref{fig:masani} and ~\ref{fig:centers}) shows independent consistency with the 5.2 kpc adopted. This was further validated by performing a fit in which the distance was varied, showing a strong preference for the value adopted. This is not a trivial result given the different distance dependencies of PM and LOS velocity components and adds further support to the distance determinations in \cite{2015ApJ...812..149W}, which exploit this effect.  This value also shows agreement with the distance determinations of  the \cite{1996AJ....112.1487H} catalog (2010 edition) and the more recent ones from \cite{2021MNRAS.505.5957B}. 

\cite{van2010new} reported central values for the central and outer anisotropy parameters of $\beta_0 = 0.13 \pm 0.02$ and $\beta_{\infty} = -0.53 \pm 0.22,$ this is consistent at the $1$-$\sigma$ level with our determinations (cf. Fig~\ref{fig:postani}), although our derived profile shows a higher degree of isotropy. This has a dependence on the cluster distance, which is significantly higher in our case (and in better agreement with recent determinations). \cite{2015ApJ...803...29W} found good agreement with isotropy for the central region of $\OC$, as measured by the proper motion anisotropy ratio, while finding some deviations for the outer regions of order $\sim 20 \%$, indicating some radial anisotropy. Our proper motion data show comparable scatter, although the trend toward radial anisotropy at larger radii is less marked. In any case, the fact that we did not find strong differences when replacing our proper motion data with those of \cite{2015ApJ...803...29W} indicates that our mildly anisotropic models that are favored are also adequate in this case.

Another factor which may play a role in central mass determinations as seen, for instance, in \cite{van2010new}, would be the photometric profile determinations. These authors found that different assumptions about the morphology of the photometric component led to differences in the degree to which central masses were favored. This is indicative of the importance of using a photometric profile which accurately reproduces the photometry. In this regard, the introduction of the $\alpha \beta \gamma$ profile as applied to this study, which was found to produce excellent fits to the photometry, constitutes a significant methodological improvement with respect to other modeling approaches which may nonetheless work well for other cases.\footnote{One such example we considered was a generic three-component Plummer model, which we found to produce significantly inferior fits to the photometry, despite having a larger number of free parameters.}

To account for the presence of rotation, we followed section 5.4.3 of \cite{2016ApJ...821...44F} by using full second velocity moments, which incorporate potential non-vanishing mean velocities acting as quadrature terms, rather than pure velocity dispersions.

As an indicative assessment of the potential effects of axisymmetry in our models, we also performed an analysis comparing with the results of \cite{Evans:2021bsh} where HST stellar kinematics from \cite{2015ApJ...803...29W} were fit using JAM axisymmetric models \citep{2008MNRAS.390...71C}, which also include a treatment of rotation. Using this dataset and adopting a Gaussian component for the dark mass (as in Eq.~\eqref{eq:deng}) and a constant anisotropy as in \cite{Evans:2021bsh}.\footnote{Note that the definition of anisotropy is not necessarily equivalent between axisymmetric and spherical models (cf. Eq.~\ref{eq:beta} with Equation 3 of \cite{Evans:2021bsh}), nonetheless, for a system that is close to isotropic and spherical this should be approximately the case, while also allowing for differences in these definitions to be reabsorbed as different values for the anisotropy profiles.} After making these adjustments with the mass modeling and data used (which excludes the stellar kinematic and pulsar timing data in our main analysis), we were able to recover the central values for both the mass of the dark and photometric components of \cite{Evans:2021bsh} within $\sim 15 \%$ accuracy when comparing with our central values.  This suggests a limited effect which may be attributable to axisymmetry and rotation, and is conservative in that these data did not include rotation signatures the way we accounted for them in our main analysis, and it does not account also for potential differences due to the photometric profiles adopted which may not be intrinsic to these effects.\footnote{We did, however, perform an earlier analysis based on the independent dataset of \cite{1995AJ....109..218T} for the photometry, finding consistency with the main results presented in this work.} This is instructive to compare with \cite{2013MNRAS.429.1887D}, who concluded that flattened models with rotation led to mass estimates for the cluster that were $10\%$ lower, with a $4.2\%$ difference in the estimate being attributable to rotation. We also note the discussion of \cite{van2010new}, where it is argued that the effects of flattening and rotation should be limited in the central region of interest.
Thus, while we intend to explore axisymmetric models in future works, we expect these to have only a moderate quantitative effect for the purposes of this study.

We have also considered the effects of different mass parametrizations for our dark component, finding that the essential conclusions of our analysis remain unchanged. This is explored in Appendix~\ref{app:diffdark}. 

Lastly, we also performed an additional analysis where we allow for a potential underestimation of errors in the LOS velocity dispersion profile. This is to account for effects such as increased shot noise, which, despite being explicitly considered, for instance, in \citet{2010ApJ...719L..60N}, may in some cases be prone to being underestimated. This was included in the form of a free parameter that corresponds to a fractional error relative to the centrally observed value which was added in quadrature when computing the likelihood. This had little effect on the central values, also favoring a central dark component over an IMBH. The main difference in this case was that some degeneracy between the concentrated dark component and the more extended one from the pulsars was present, meaning that the data could not fully discriminate between these. While the inclusion of LOS velocities is important in constraining our mass models, this further attests to the robustness of our results, even if a significant underestimation of errors were to be present. 
An analysis with additional datasets not originally included in this work will also be performed in a follow-up study.



\subsection{A dark cluster of remnants in $\OC$}
\label{sec:rem}

It is apparent from our results in Sec.~\ref{sec:fit} that the two kinematically relevant components, namely, the extended central mass and photometric profiles, are very well constrained by the data. In this context, it is instructive to compare our results with the analyses of \cite{2019MNRAS.482.4713Z, baumgardt2019no, 2024MNRAS.529..331D}, where the presence of a cluster of stellar-mass black holes is determined to be a viable alternative to the IMBH hypothesis to account for the observed velocity dispersions in $\OC$. Stellar-mass black holes also arise naturally in dynamical simulations, where they have been found to play an important role in the core dynamics of GC analogs, and are predicted to be hosted in significant numbers by many present-day GCs \citep{2020IAUS..351..357K}. Indeed, a concentrated cluster of remnants appears to be the only viable interpretation for the dominant mass component favored for the central region that our analysis favors over any of the other mass components considered. A two-component model of light and heavier remnants as the kinematically relevant contributions is indeed favored by our model, in agreement with the theoretically motivated assumptions of these studies. 

\cite{2024MNRAS.529..331D} performed a recent analysis of $\OC$ among other GCs, in which {{\tt limepy}} distribution function models \citep{2015MNRAS.454..576G} were used to obtain realistic distributions for an ample spectrum of species comprising stellar-mass black holes, white dwarfs, neutron stars and main sequence stars. These were subsequently fitted to velocity dispersion and stellar mass function data following the procedure described in \cite{2023MNRAS.522.5320D}. 
A key finding from this study is the presence of a substantial central mass component dominated by stellar-mass black holes, with total mass $1.82^{+0.05}_{-0.06} \times 10^5 \: \MS,$ while the remaining lighter remnants follow the same profile to a good approximation. This favors a two-component mass model whereby the segregation of lighter objects is inhibited. This shows broad agreement with previous studies of $\OC$ where this has been argued, owing to its young dynamical age, long two-body relaxation time, and black hole population abundance, which has been found to be anti-correlated with mass segregation \citep{2024MNRAS.529..331D}. We were able to explicitly compare our results with the projected mass profiles from \cite{2024MNRAS.529..331D}.  We found that for the region beyond $\sim 1$ pc, the profile from \citep{2024MNRAS.529..331D} closely matched a Plummer sphere with scale radius $\sim 1.5$ pc,\footnote{We note that, while the modeling of \cite{2024MNRAS.529..331D} indicates a more concentrated distribution for the remnants below this radius which is not reproduced by the Plummer model, our inclusion of a point mass has the ability to compensate for this should a higher enclosed mass be favored for this region. The fact that it does not do so at a statistically significant level indicates that such feature, while physically expected from the remnant models of \cite{2024MNRAS.529..331D}, is likely not relevant for the kinematics considered in our analysis. } compared to our derived value of $1.88 \pm 0.10$ pc. While this implies a profile that is systematically more concentrated than the one we derive, it agrees at the $\sim 1\sigma$ level with the $\sim 20 \%$ less massive and extended profile we obtained without the inclusion of MSP accelerations. This result is shown in Appendix~\ref{app:noaf} (Figs.~\ref{fig:noaf_mass} and \ref{fig:noaf_pul}), where a fit without the inclusion of MSP accelerations is performed for comparison. Interestingly, this suggests the analysis of \cite{2024MNRAS.529..331D} may benefit from the inclusion of MSP accelerations following our implementation, leading to a profile that would likely be more consistent with the one we obtain. Notwithstanding systematic uncertainties due to modeling assumptions, which should be fully accounted for a more definitive comparison, the similarity of these profiles, and the fact that lighter remnants would not be able to undergo the degree of segregation implied by our derived central mass distribution, stellar-mass black holes indeed appear to be the favored interpretation for our derived central component.


\subsection{An IMBH in $\OC$?}

\label{sec:DM}

\cite{häberle2024fastmoving} recently derived a lower bound on the mass of an IMBH in $\OC$ of $8,200 \: \MS$ based on the observation of centrally located fast-moving stars, in apparent tension with our $3\sigma$ bound of $6 \times 10 ^3 \: \MS$. However, this bound is  dependent on the assumed escape velocity for the cluster, which, for instance, is $\sim 10 \%$ higher in our analysis due to the presence of an extended dark mass attributable to stellar remnants (see Sec.~\ref{sec:rem}).

\cite{2024MNRAS.528.4941P}, on the other hand, found no evidence of an IMBH in the form of fast-moving stars in the inner $\sim 0.5$\,pc region predicted by N-body IMBH simulations from \cite{2017MNRAS.464.2174B}. \cite{baumgardt2019no} arrived at a similar conclusion based on the lack of observed fast-moving stars in their analysis, concluding that a $\sim 4\times10^4\: \MS$ IMBH, favored by other analyses  \citep{Noyola:2008kt, 2017MNRAS.464.2174B}, is ruled out in favor of a cluster of remnants, which is consistent with our results.

IMBH formation simulations in clusters have also found it challenging to produce black holes exceeding $\sim 500 \: \MS$ due to the ejection from gravitational wave recoils in black hole merger interactions \citep{2023MNRAS.526..429A,2024arXiv240606772F}. Interestingly, in recent simulations from \cite{2024arXiv240606772F}, it was found to be possible to produce $\gtrsim 10^3 \: \MS$ IMBHs in GC analogs via very-massive star formation processes. However, the authors conclude that this formation scenario disallows IMBHs of masses greater than $10^4 \: \MS,$ even for the case of very massive GCs such as $\OC.$ This poses a theoretical challenge for any claims of IMBH detections in GCs exceeding this limit.


Many works have suggested that $\OC$ is the stripped remnant of a nucleated dwarf \citep{Bekki:2003qw, wirth2020formation, johnson2020most, 2024arXiv240519286G}. If so, it should have a dark matter halo and a correspondingly higher escape velocity. This could explain the fast-moving stars found in \cite{häberle2024fastmoving} without the need for such a massive IMBH,\footnote{Note that this does not a priori imply the absence of an IMBH, as our derived bounds are still compatible with lighter IMBHs.}  or also the observation from \cite{2024MNRAS.528.4941P} of a fast-moving star with a high (0.99) cluster membership probability that appears too distant from the center ($\gtrsim 1.5$ pc) to be accounted for by an IMBH. To assess the viability of this explanation, more work is needed, including comparison with simulations and data not originally included in this analysis. This is, however, beyond the scope of this work. We will test this idea quantitatively in forthcoming papers in which we will add a dark matter halo to our $\OC$ models.

\section{Conclusion}
\label{sec:conclusion}
The results and main findings of this work can be summarized as follows: 
\begin{itemize}
\item We performed a combined analysis of stellar kinematics and LOS accelerations of MSPs to probe the mass contents of $\OC$. We explored the existence of a central dark mass component and considered competing interpretations of its nature: a cluster of stellar remnants or an IMBH. We also exploited the data to model and constrain the MSP distribution, using this to probe MSP formation models, and as an additional mass component in the model that traces stellar remnants of intermediate mass. 
\\
\item There are two key results from our analysis which are not present in previous works: One is that a two-component model with a significantly extended central mass distribution (with scale radius between $1.5$ and $2.2$ pc at the 3$\sigma$ level) is strongly favored over both an IMBH (or more concentrated distributions) and the more extended component traced by the MSPs. The other result,  which is a consequence of the previous one, is that we arrive at stringent bounds on the kinematic contributions of the two additional components considered, establishing a coexistence region between the favored central mass distribution and a putative IMBH with a $6\times 10^3\: \MS$ 3$\sigma$ CL upper limit on its mass. This altogether favors a dark cluster of remnants, rather than an IMBH, as the explanation for the central kinematics of $\OC$.
\\
\item While the MSP profile, with a $3\sigma$ CL upper bound of $6 \times 10^4 \: \MS,$ was found not to contribute meaningfully to the dynamical mass distribution of $\OC,$ we found that the MSP timing constraints do play a significant role in the kinematics, favoring a central distribution that is $\sim 20 \%$ more massive and extended. 
\\
\item We found that the MSP density profile goes as: $$\rho_{p}(r) \propto \rho_{\star}(r)^\gamma / \sigma(r),$$ with $\gamma = 1.9 \pm 0.3,$ consistent with models in which MSPs originate from stellar encounters. It is a natural expectation from this scenario that $\gamma \sim 2.$ We found, further, that the stars and X-ray sources in $\OC$ are more radially extended than the MSPs and follow a similar density profile to one another. Our analysis leads to a significant novelty: that the encounter rate-MSP abundance correlation can be extended at the intra-cluster level, motivating a profile for the spatial distribution of these objects.
\\
\item In future work, we will explore the presence of a dark matter halo in our mass model, as is predicted to be present if $\OC$ is the dissolved remnant of an accreted, nucleated, dwarf \citep[e.g.][]{2024arXiv240519286G}.
\\
 \item Despite limited observations leading to full timing solutions for only 5 of the 18 recently discovered MSPs in $\OC$, and more potential discoveries still underway \citep{Chen:2023lzp}, our analysis shows the potential and promise of combining velocity dispersion data with pulsar accelerations to probe the inner mass distribution of GCs. Future observations may allow probing whether extremal accelerations occur for central MSPs, a `smoking-gun' signature of an IMBH and a disfavored scenario for $\OC$ in our analysis. 
 \\ 
 \item  With the number of MSP discoveries in GCs having recently doubled over a 5-year period amid the advent of radio surveys with unprecedented sensitivity, such as MeerKAT, FAST, and notably the upcoming SKA, it is foreseeable that the methodology we introduce will be increasingly relevant for the understanding of the kinematics of these systems.
 \end{itemize}


\begin{acknowledgements} 

We thank the anonymous referee for very useful comments and feedback that have improved this paper.
We are grateful to Renuka Pechetti for kindly providing us the photometric data used in this work.
We would like to thank Vincent Hénault-Brunet, Giuseppina Battaglia, José María Arroyo Polonio, Paul Beck, Addy J. Evans, Jorge Sánchez Almeida, Claire S. Ye, and P.C.C. Freire for instructive discussions and their interest in our work. We also thank Joshua S. Speagle for providing us valuable information about the use of \texttt{dynesty}.
We are grateful to Jorge Terol Calvo, Jorge García Farieta, and Ángel de Vicente for the invaluable support provided in using the high-performance computing systems at Insituto de Astrofísica de Canarias (IAC).
We are grateful to Elena Pinetti and the team at the Cosmic Physics Center at Fermilab, including Alex Drlica-Wagner and Albert Stebbins, where this work was presented and discussed. 
The authors wish to acknowledge the contribution of the IAC High-Performance Computing support team and hardware facilities to the results of this research. AB, FC and JMC acknowledge funding received from the European Union through the grant ``UNDARK'' of the Widening participation and spreading excellence programme (project number 101159929).
AB and JMC acknowledge support from the MICINN through
the grant “DarkMaps” PID2022-142142NB-I00. JIR would like to acknowledge support from STFC grants ST/Y002865/1 and ST/Y002857/1.
This work has made use of the following software packages: \texttt{dynesty} \citep{2020MNRAS.493.3132S, 2022zndo...6609296K}, \texttt{GravSphere} \citep{2017MNRAS.471.4541R, 2018MNRAS.481..860R, 2020MNRAS.498..144G, 2021MNRAS.505.5686C}, \texttt{corner.py} \citep{corner}, \texttt{NumPy} \citep{Harris_2020}, \texttt{SciPy} \citep{2020SciPy-NMeth}, \texttt{Matplotlib} \citep{Hunter:2007}, \texttt{Jupyter Notebook} \citep{2016ppap.book...87K}. 

\end{acknowledgements}

\bibliographystyle{aa} 
\bibliography{references}

\begin{appendix}

\section{Posterior distributions of the anisotropy parameters}
\label{app:postani}

\begin{figure}[!htb]
     \centering
 \includegraphics[width=\linewidth, keepaspectratio]{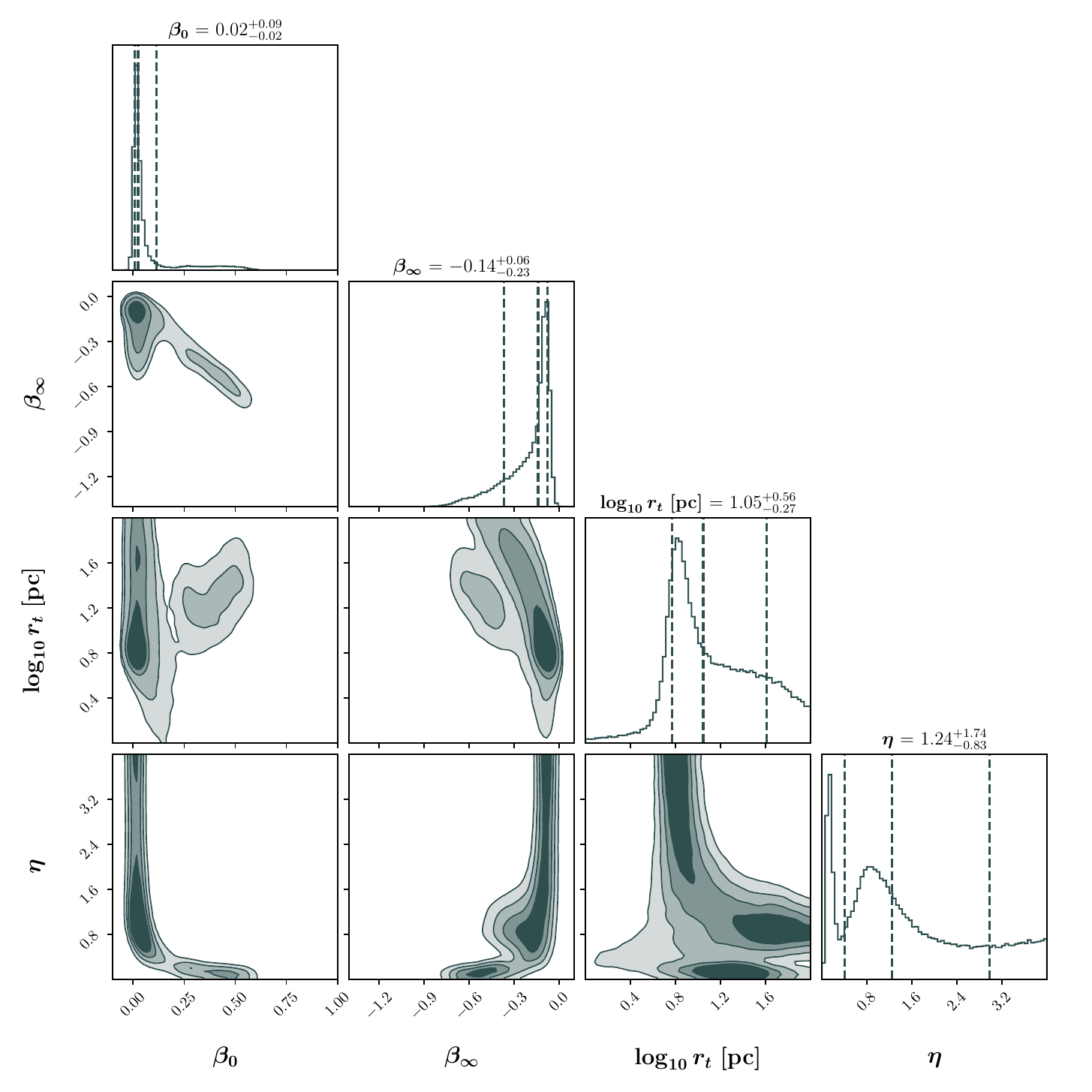}
 \caption{Posterior distributions of the anisotropy parameters. The parameter space exhibits multiple degeneracies, but due to the correlations inherent to these, the resultant anisotropy profile is well constrained (cf.~Fig.~\ref{fig:masani}). Note that this is not inconsistent with our results and that these degeneracies are the result of our parametric modeling and not intrinsic to the anisotropy profile. 
 \label{fig:postani}}
\end{figure}


\section{Effects of excluding MSP accelerations}
\label{app:noaf}

\begin{figure}[!htb]
     \centering
 \includegraphics[width=\linewidth, keepaspectratio]{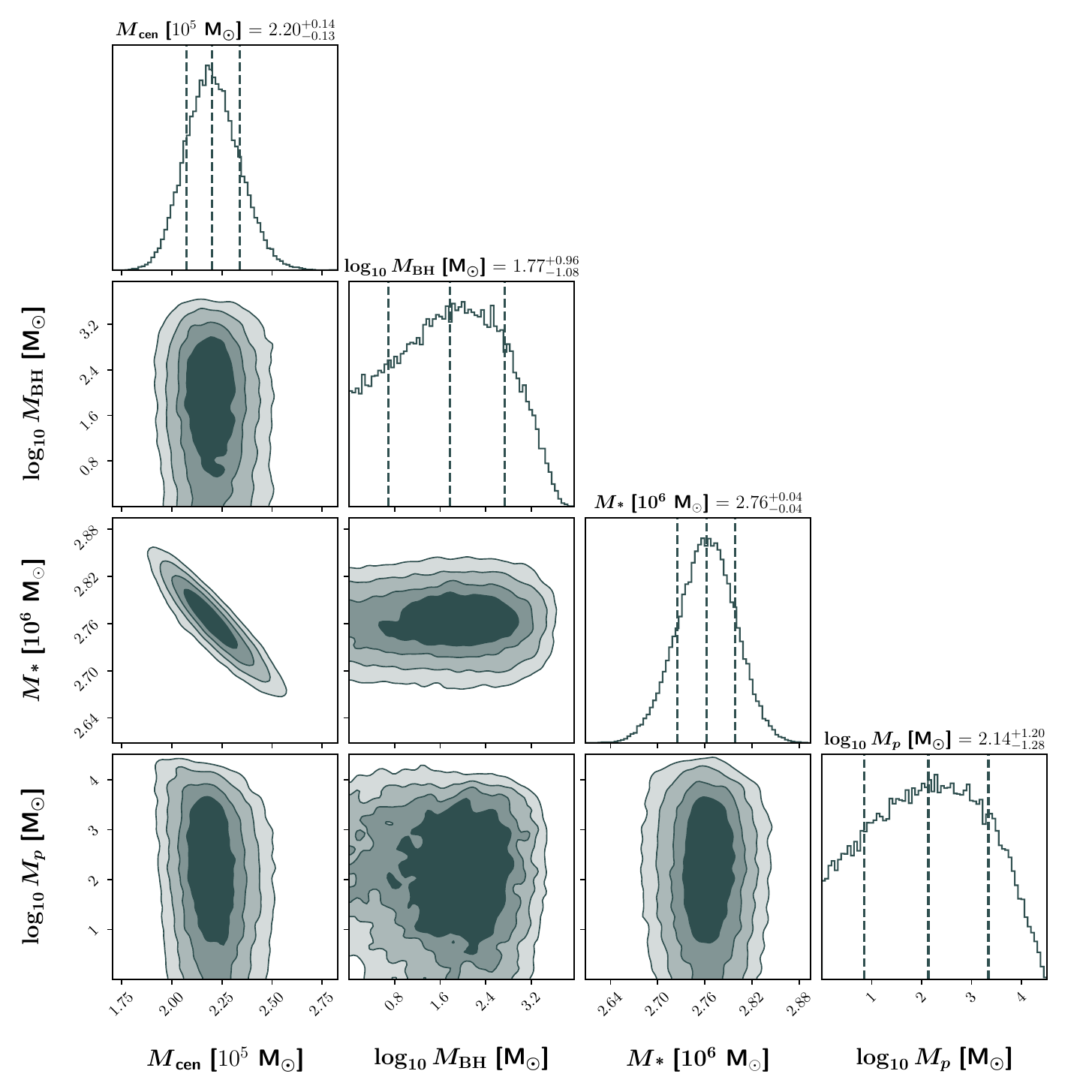}
 \caption{Posterior distributions for the masses of the components considered during a fit performed without the inclusion of MSP LOS accelerations into the likelihood function. The median for the central mass from the full fit (cf. Fig.~\ref{fig:masani}) is $ \sim 20 \%$ greater than the one found during this fit, with the other components showing only moderate differences.
 \label{fig:noaf_mass}}
\end{figure}

\begin{figure}[!htb]
     \centering
 \includegraphics[width=\linewidth, keepaspectratio]{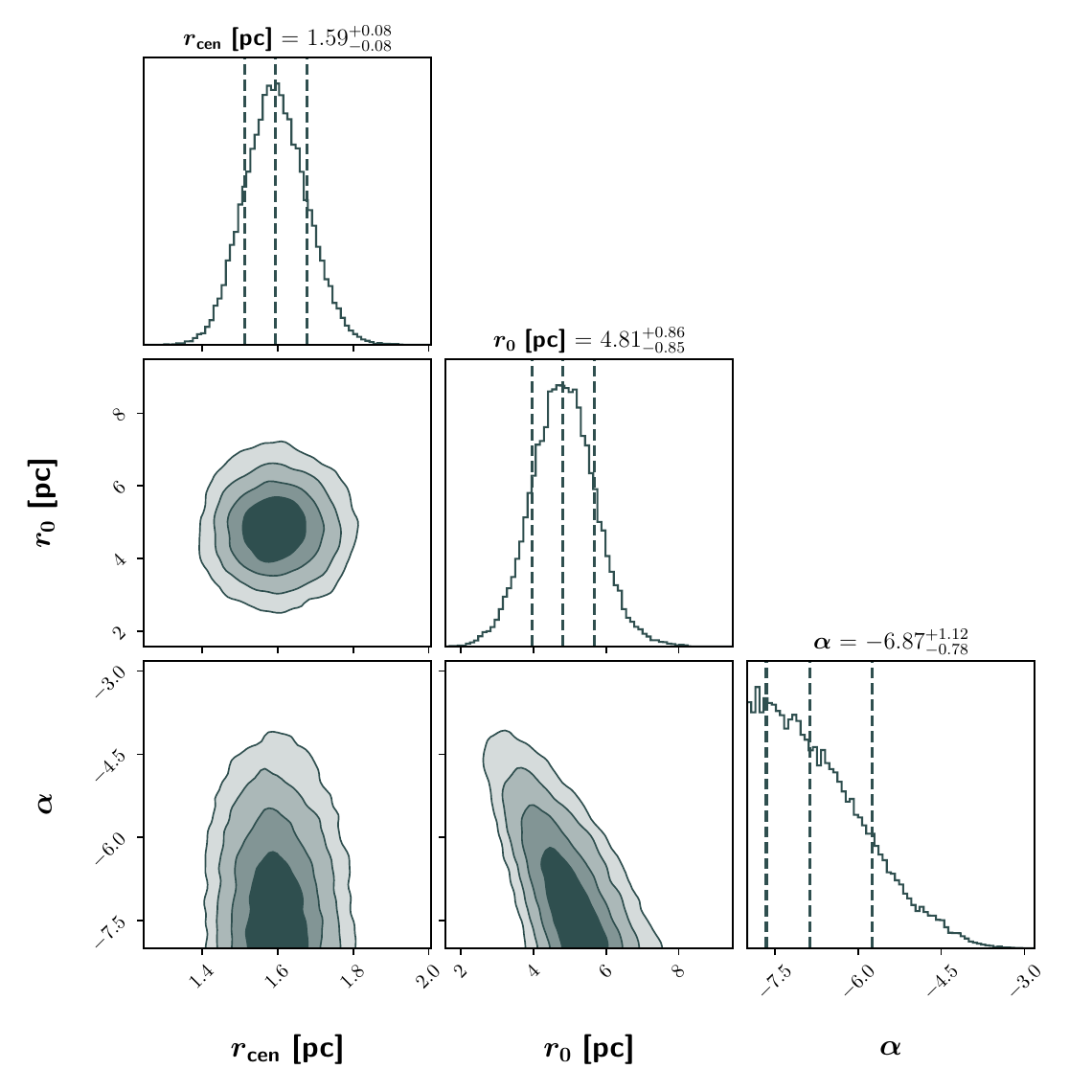}
 \caption{Posterior distributions for the morphological parameters of the mass profiles without the inclusion of MSP LOS accelerations into the likelihood function. 
 The median for the central mass length scale from the full fit (cf. Fig.~\ref{fig:masani}) is $\sim 18\%$ greater than the one found during this fit.
 The morphology of the MSP distribution is still constrained due to the projected radii included in the positional component of the likelihood function, favoring a somewhat different posterior distribution when accelerations are excluded. This, however, has a negligible effect on the remainder of the mass components in the absence of accelerations to constrain their kinematics. 
 \label{fig:noaf_pul}}
\end{figure}
\section{Use of different centers}
\label{app:centers}

\begin{figure*}[h!]
    \centering
\includegraphics[width=7.5cm]{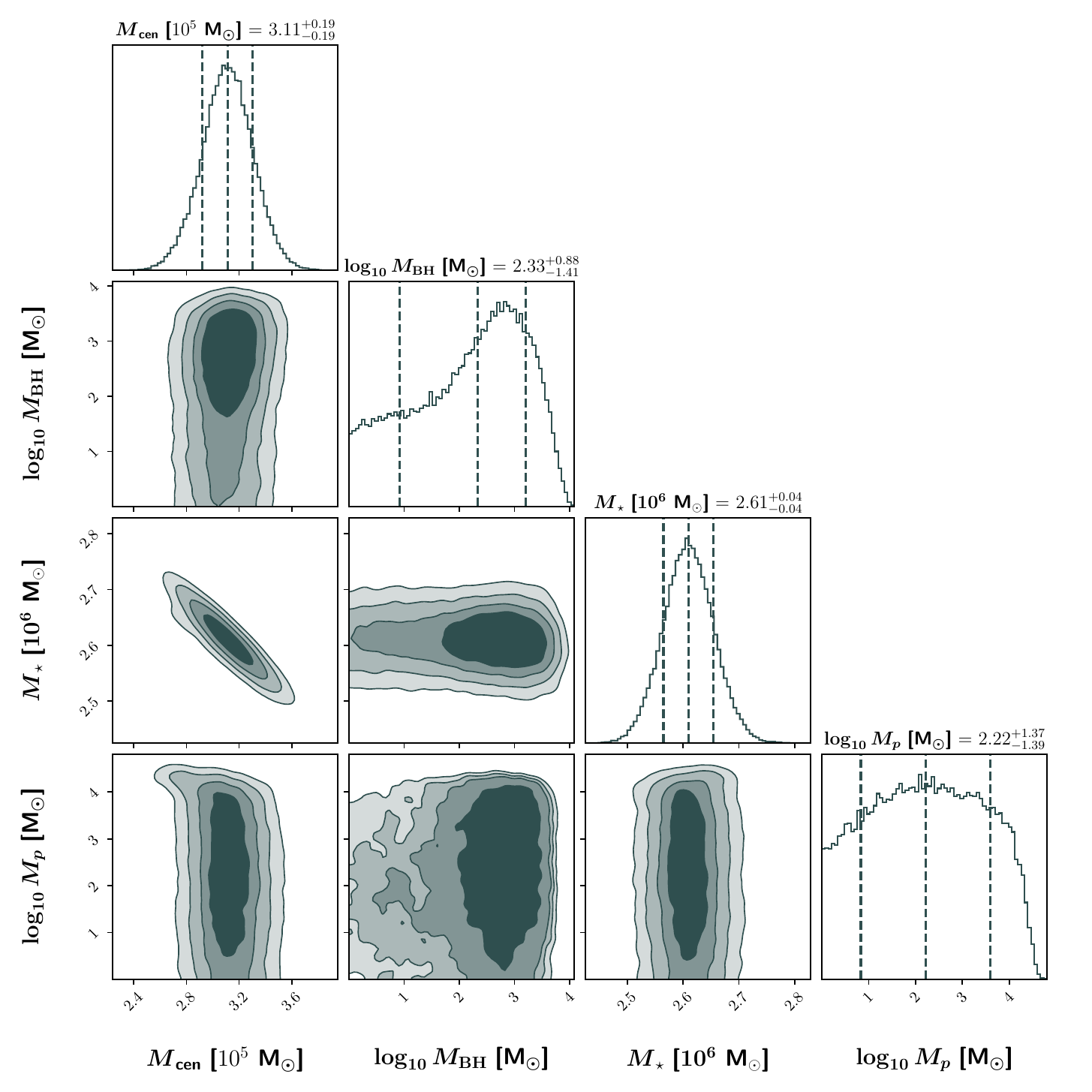} 
\hspace{1.5cm}
\includegraphics[width=7.5cm]{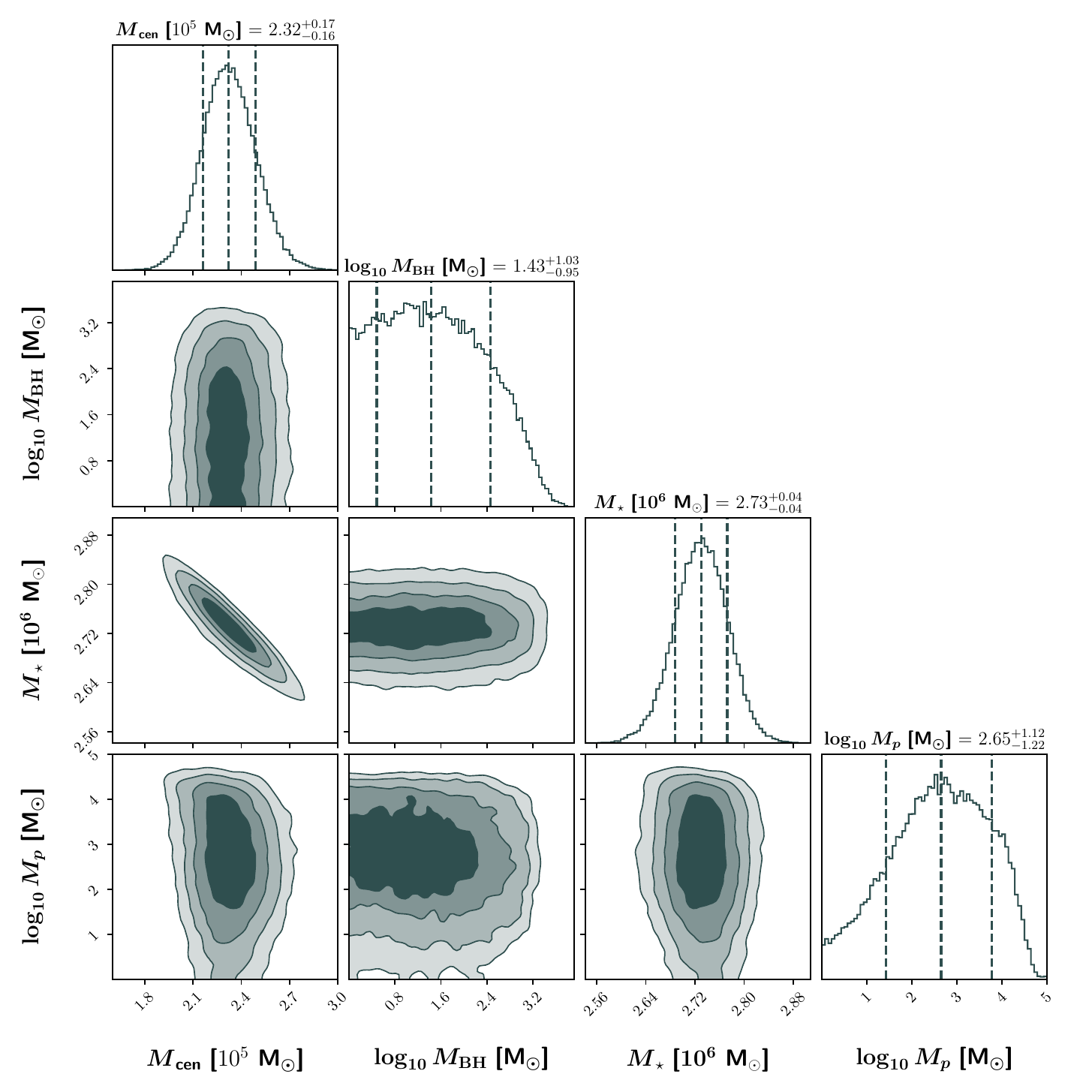}
\includegraphics[width=7.5cm]{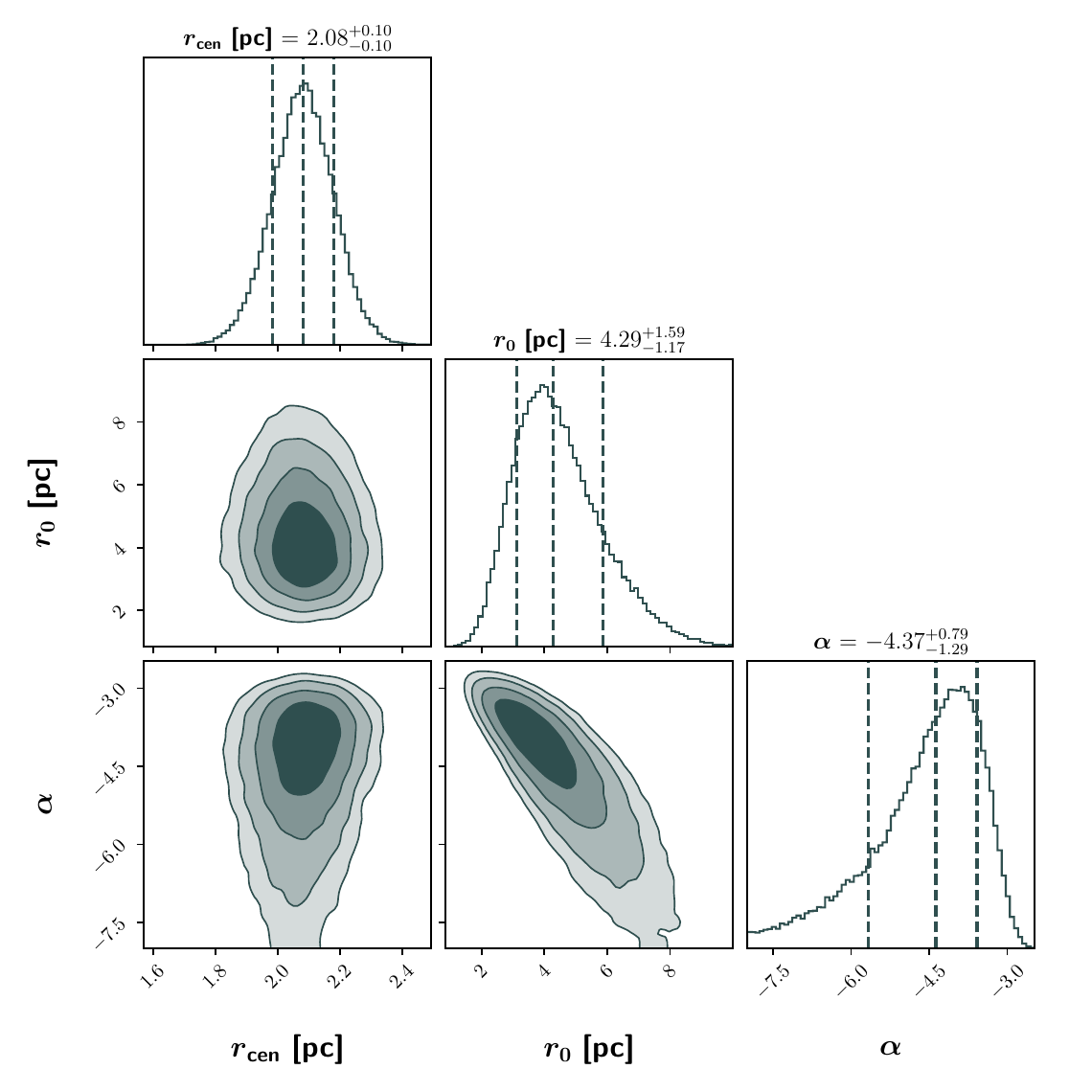} 
\hspace{1.5cm}
\includegraphics[width=7.5cm]{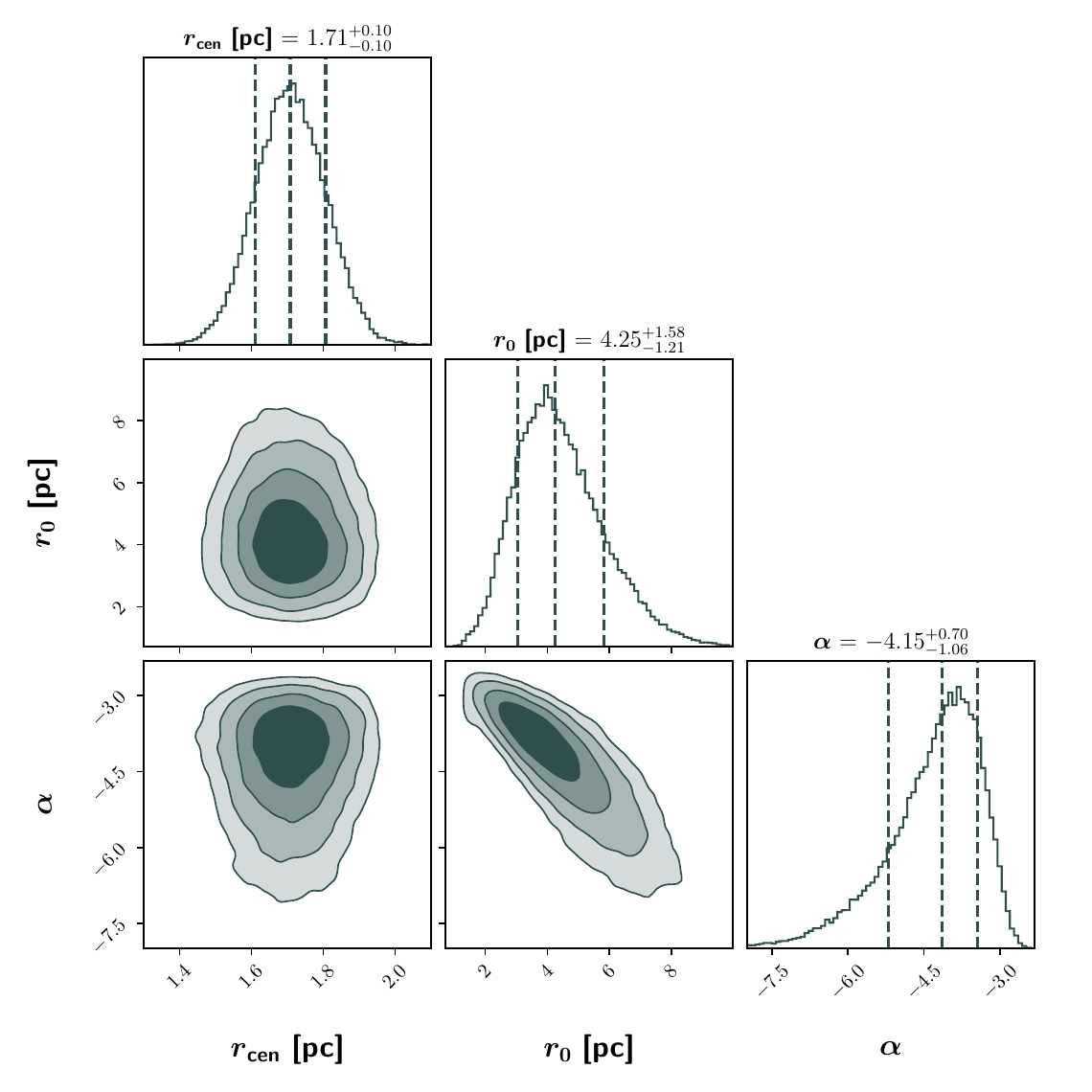}
 \caption{
 \emph{Upper-left:} Posterior distributions for the mass components considered in our kinematic models using the Noy10 center. The median of the central mass component for the And center, is $\sim 15 \%$ lower, consistent with a more extended central mass distribution for the Noy10 center. The stellar mass obtained is only marginally smaller ($\lesssim 3 \%$), with the other IMBH and MSP components remaining subdominant without significant statistically meaningful differences.
   \emph{Upper-right:} Same as the upper-left plot, but for the Noy08 center. In this case, the central mass for our main analysis with the And center is $\sim 14 \%$ higher than the result shown, consistent with a more concentrated distribution for the Noy08 center, although still with the $1 \sigma$ CL regions overlapping each other. A small increase of $\lesssim 2 \%$ is observed with respect to the stellar mass value, but these central values lie within their respective $1 \sigma$ CL regions. The IMBH and MSP mass components, once more, show approximate consistency with the previous results.$^\dagger$
   \emph{Lower-left:} Posterior distributions of the morphological parameters for the mass models using the Noy10 center. The median of the scale radius using the And center is $\sim 11 \%$ lower, indicating a somewhat more extended central mass distribution for the Noy10 center, with the $1 \sigma$ CL regions barely overlapping each other. The MSP parameters agree at the $1 \sigma$ level.
    \emph{Lower-right:} Same as the middle-left plot, but for the Noy08 center. In this case, the median of the scale radius using the And center is $\sim 10 \%$ greater, indicating a somewhat more concentrated central mass distribution, but still showing $1 \sigma$ CL compatibility. As with the Noy10 center, the MSP parameters show $1 \sigma$ level consistency with the And center results. 
    \\
    $^\dagger$ Marginal differences in the morphology of the posterior distributions of the IMBH and MSP masses are not necessarily statistically meaningful, owing to the fact that, due to being inherently less constrained, these may require additional independent runs to fully establish such differences (which we did perform for the case of the And center). This, however, is not necessary for the purposes of the complementary analysis presented in this appendix.
\label{fig:post_centers}}
\end{figure*}

\begin{figure*}[h!]
    \centering
\includegraphics[width=6.83cm]{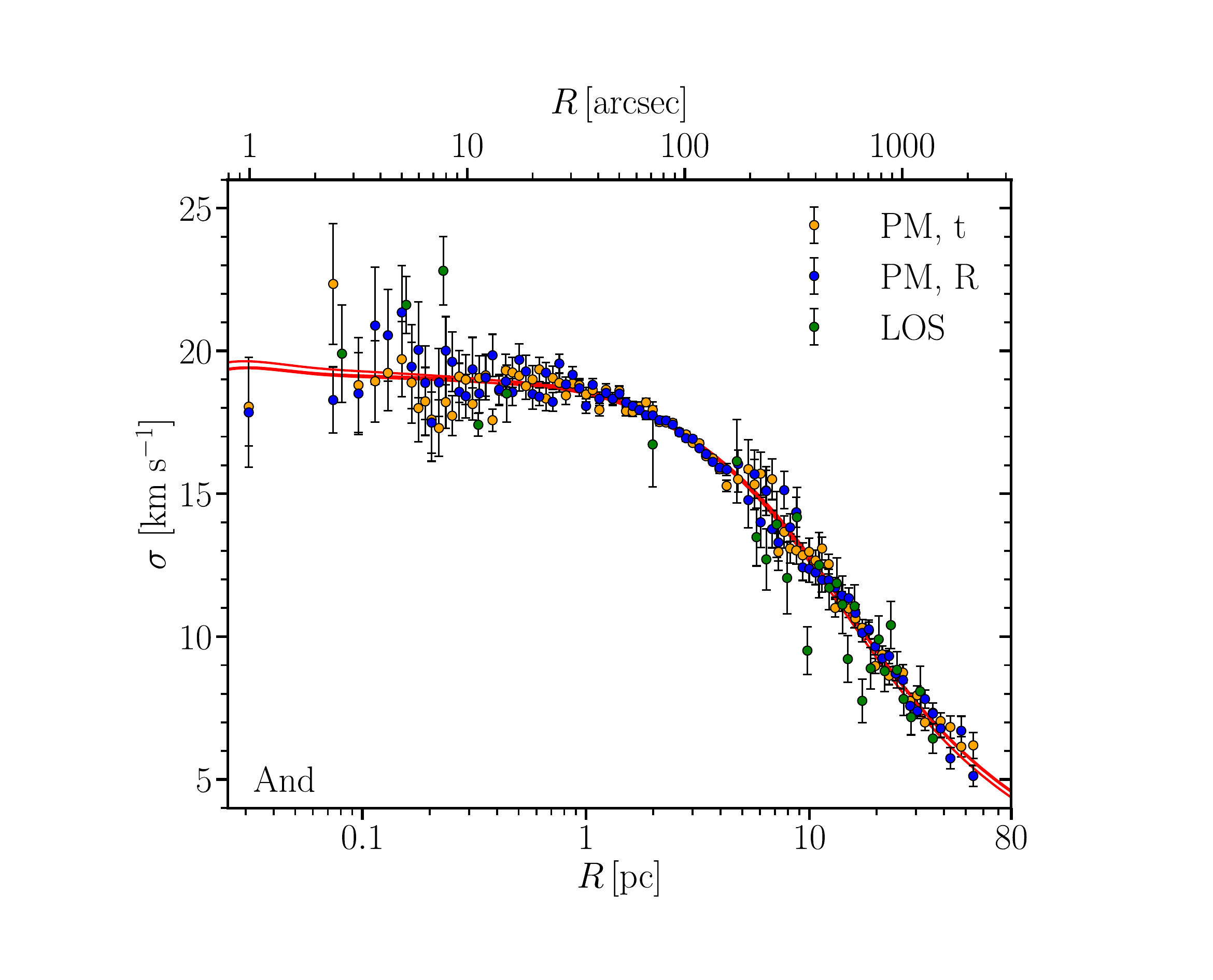}
\hspace{-1.21cm}
\includegraphics[width=6.83cm]{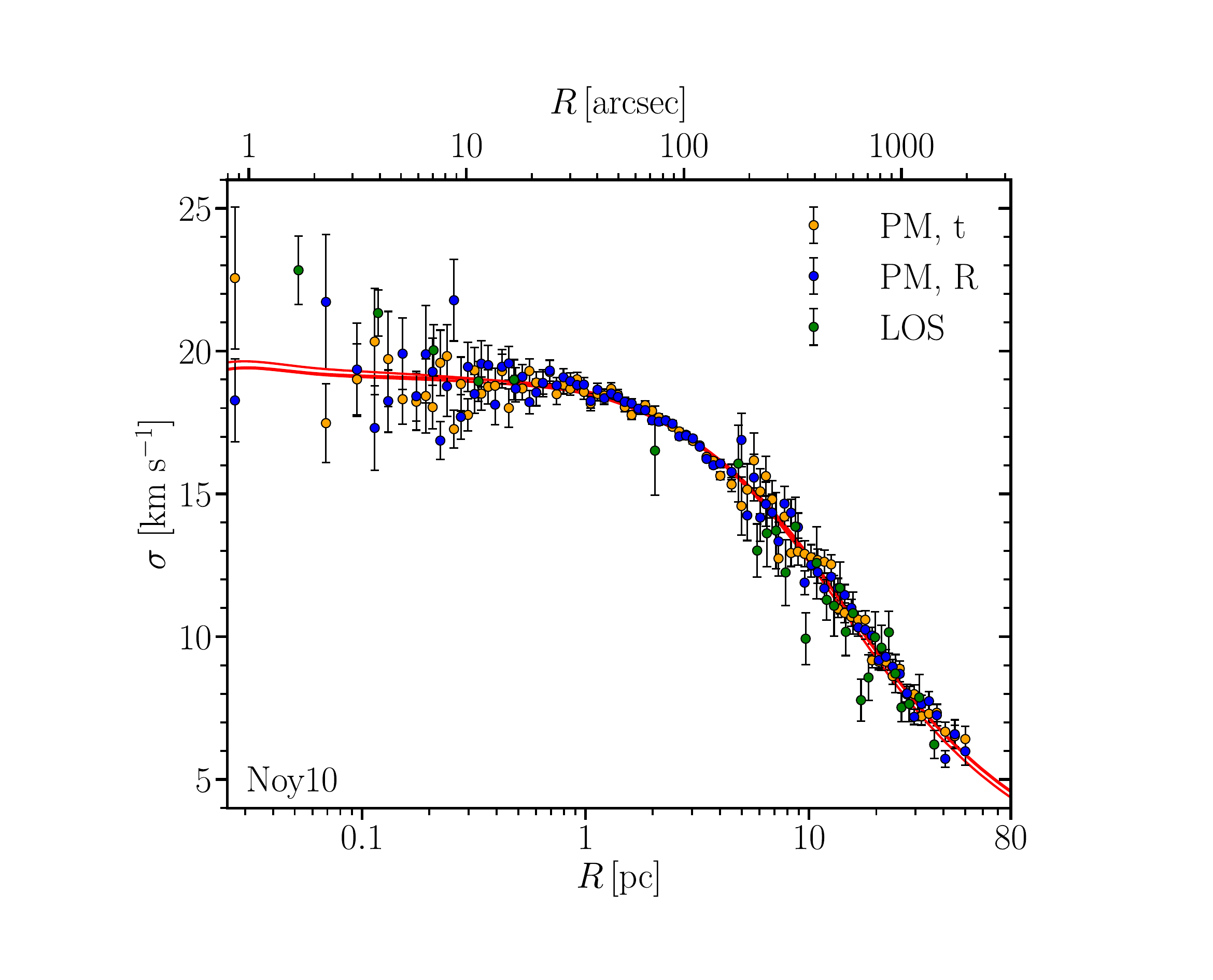}
\hspace{-1.21cm}
\includegraphics[width=6.83cm]{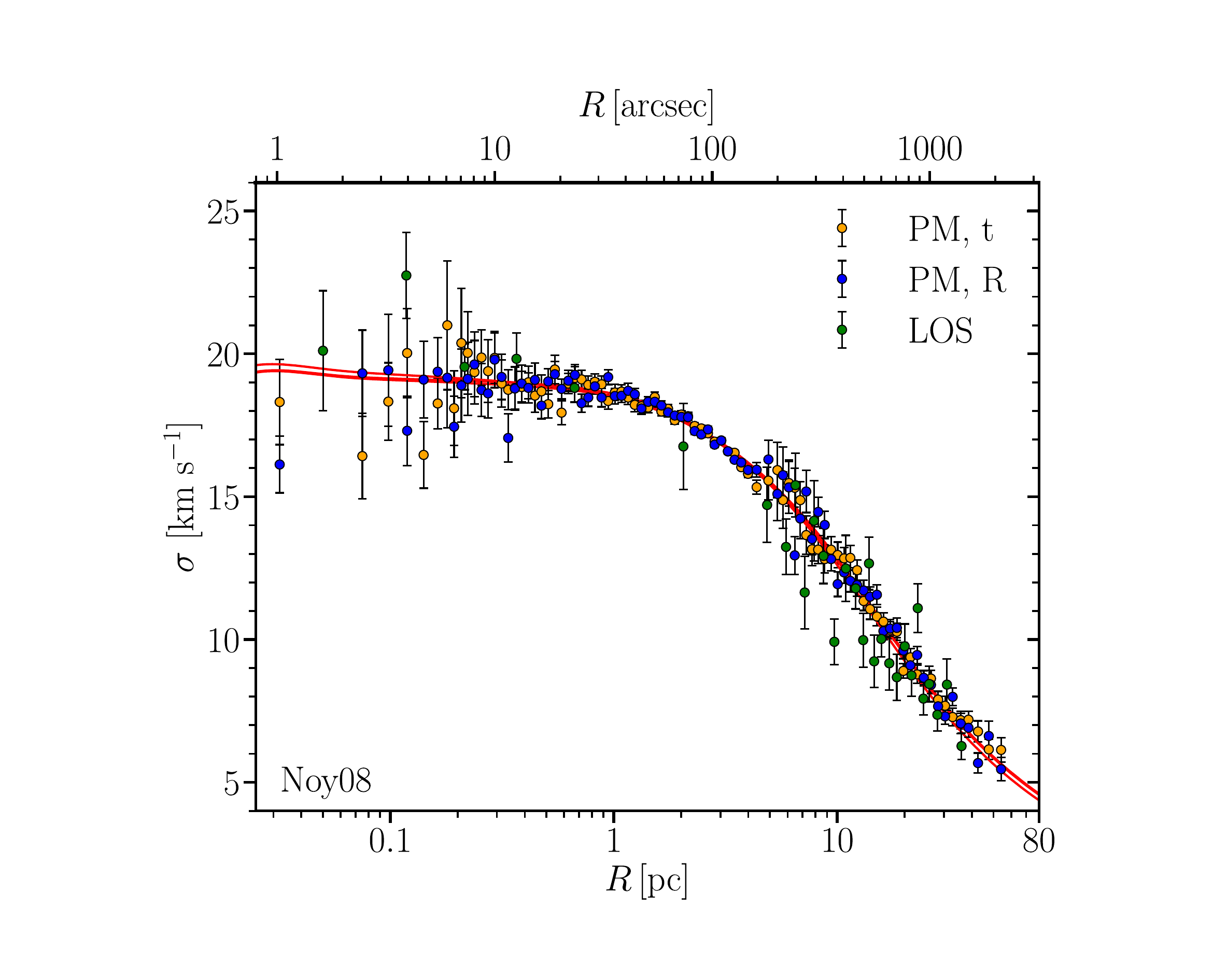}
 \caption{Combined velocity dispersion profiles for the three components using the \cite{anderson2010new} center (And, \emph{left})  \cite{2010ApJ...719L..60N} (Noy10, \emph{middle}) and \cite{Noyola:2008kt} (Noy08, \emph{right}) kinematic centers. The maximum-posterior velocity dispersion profile (red) from our main analysis using the And center is shown for reference in all plots, yielding fits of comparable quality for all the cases considered. Due to the close-to-isotropic behavior of our inferred distribution at the 5.2 kpc distance employed, the three components of our maximum posterior fits are almost identical, with the large majority of the data points showing close overlap with each other. 
\label{fig:centers}}
\end{figure*}
 



\section{Use of different dark components}
\label{app:diffdark}

\begin{figure}[!htb]
     \centering
 \includegraphics[width=\linewidth, keepaspectratio]{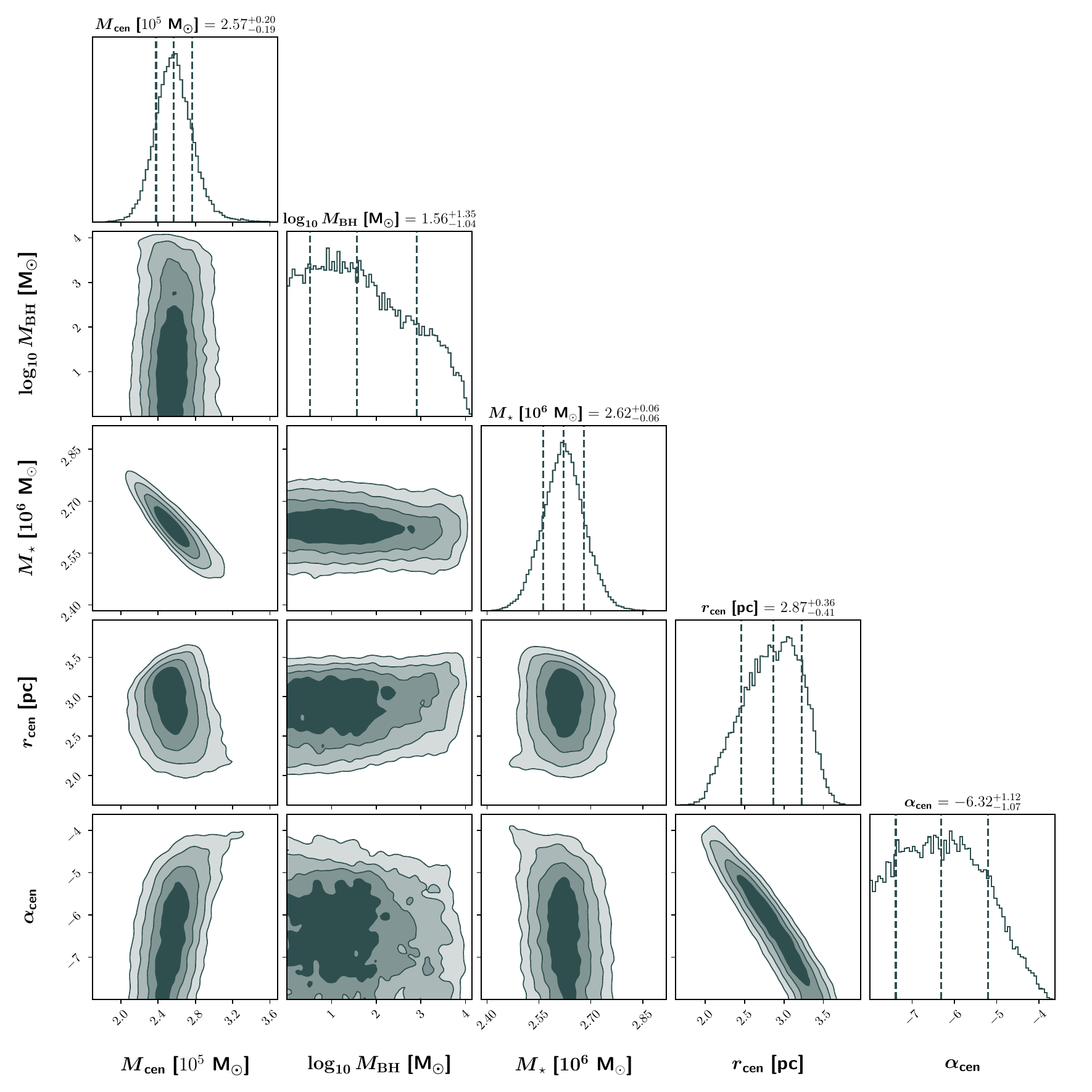}
 \caption{Posterior distributions for the mass model parameters using the generalized mass profile in Eq.~\eqref{eq:massmsp} for the extended dark component.
 \label{fig:bhgcext}}
\end{figure}

\begin{figure}[!htb]
     \centering
 \includegraphics[width=\linewidth, keepaspectratio]{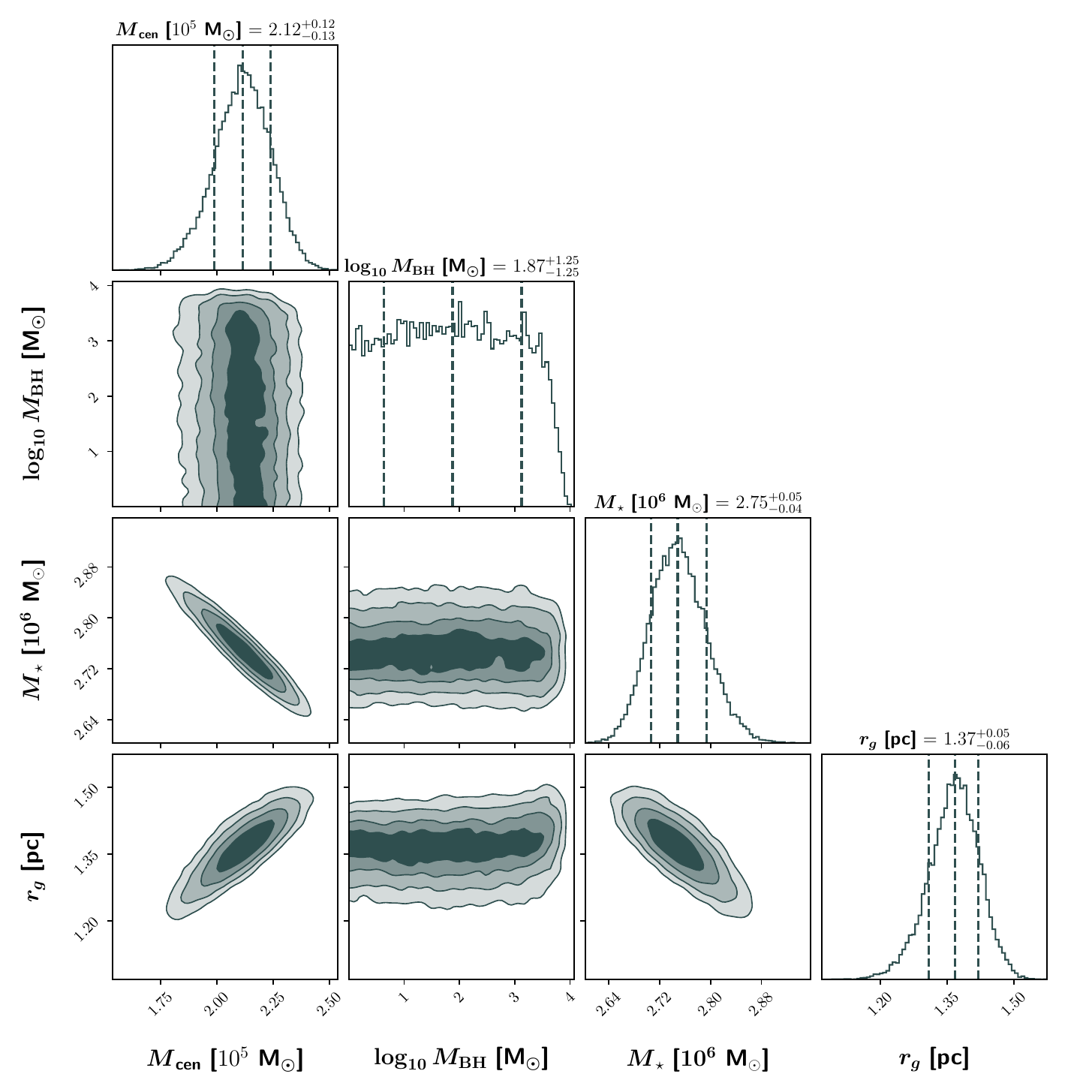}
 \caption{Posterior distributions for the mass model parameters using a Gaussian profile (Eq.~\eqref{eq:deng}) for the extended dark component.
 \label{fig:gcext}}
\end{figure}

Figs.~\ref{fig:bhgcext} and \ref{fig:gcext} explore the influence of differences in the adopted mass models for the extended dark component. In the first case, we adopt the functional form of the generalized profile in Eq.~\eqref{eq:massmsp} with unconstrained priors (i.e. not ascribing to it constraints from the pulsar distribution as in our main analysis). This has the advantage of having a great deal of flexibility, as the exponent parameter $\alpha$ reproduces distributions of varying concentration, including the Plummer and King model profiles, and the parameter space explored allows anything from something approximating a point mass to a distribution of greater extension than the photometric profile. For both of these fits, we have left the point mass component due to a putative IMBH unchanged. In this case, we find that the distribution of all the mass components are compatible at the $1 \sigma$ level, with an IMBH still being disfavored as a significant contributor to the kinematics. Note that the scale radius (which is mathematically identical to $r_0$ in Eq.~\eqref{eq:massmsp}) may appear to indicate a more extended distribution when compared to our earlier result with the Plummer model, but this is due to a degeneracy in the $r_{\rm cen} - \alpha_{\rm cen}$ plane which is fixed in the former case and has little physical impact on our results. This result further corroborates that a) the favored dark component in our analysis is robust with respect to more general profile parametrizations that have the ability to reproduce it and b) that an additional extended distribution traced by the pulsars is not favored by these observations (consistent with the expected level of mass segregation in $\OC$, both theoretically and observationally). 

For the second case we adopt the density of a spherical Gaussian profile as in \cite{Evans:2021bsh}, which can be expressed as:
\begin{equation}
\label{eq:deng}
    \rho_{\rm cen, \: g}(r) = \frac{M_{\rm cen}}{(2 \pi r^2_{g})^{3/2}} \exp \Bigg( -\frac{r^2}{2 r^2_{\rm g}}  
    \Bigg).
\end{equation}
In this case, we observe a dark component that is systematically more concentrated, with a mass that is $\sim 20 \%$ lower in this case, while the photometric profile is somewhat more massive, but with the $1 \sigma$ intervals of the posteriors still overlapping. This, however, does not change the essential conclusions of our analysis of these distributions, namely, the presence of an extended dark component of $\sim 2 - 3 \times 10^5 \: \MS$ and the absence of a kinematically relevant IMBH component, capped at a few thousand solar masses.
\end{appendix}
\end{document}